%% file: PDK_lp2pi0.tex
\documentclass[%
 reprint,
amsmath,amssymb,
 aps,
 prd,
 superscriptaddress,
 reprint,
 longbibliography
floatfix,
]{revtex4-2}

\usepackage[utf8]{inputenc}
\DeclareUnicodeCharacter{2032}{$^\prime$}
\usepackage{graphicx}
\usepackage{dcolumn}
\usepackage{bm}
\usepackage{romannum}
\usepackage{booktabs}
\usepackage{tabularx}
\usepackage{multirow}
\usepackage{filecontents}
\usepackage{makecell}
\usepackage{afterpage}
\usepackage{etoolbox}
\usepackage{hyperref}
\usepackage{placeins}
\usepackage{enumitem}
\usepackage{orcidlink}
\let\oldauthor\author
\renewcommand{\author}[2][]{%
  \if\relax\detokenize{#1}\relax
    \oldauthor{#2}%
  \else
    \oldauthor{#2\,\orcidlink{#1}}%
  \fi
}

\makeatletter
\providecommand{\href@noop}[0]{\@secondoftwo}
\makeatother

\begin{document}

\preprint{APS/123-QED}

\title{Search for proton decay via $p\rightarrow e^+\pi^0\pi^0$ and $p\rightarrow \mu^+\pi^0\pi^0$ in 0.401 megaton-years \\exposure of Super-Kamiokande I-V}

\input{authors_20230602_20260407-orcid.tex}
\date{\today}
\begin{abstract}
We searched for proton decay via $p\rightarrow e^+\pi^0\pi^0$ and $p\rightarrow \mu^+\pi^0\pi^0$ in 0.401 megaton-years of data collected in all pure water detector phases of Super-Kamiokande (SK) I-V. 
A theoretical study predicts proton decay rates without assuming a particular grand unified theory and suggests that three-body proton decays involving two pions can have decay rates comparable to those of $p\rightarrow e^+\pi^0$ and $p\rightarrow \mu^+\pi^0$.  
This is the first search for proton decay into a charged anti-lepton and two neutral pions in SK. 
One data candidate event was found for each of the two decay modes, which is compatible with the expected atmospheric neutrino background. 
We set lower limits on the lifetime of $\tau/B(p\rightarrow e^+\pi^0\pi^0 ) > 7.2\times10^{33}$  years and $\tau/B(p\rightarrow \mu^+ \pi^0\pi^0) > 4.5\times10^{33}$ years at 90\% confidence level. 
These limits are more than one order of magnitude better than those of the previous experiment.

\end{abstract}
\maketitle


\section{\label{sec:lab1}Introduction}

 Grand Unification Theories (GUTs)~\cite{pdgGUT2023, GeorgiSU(5), LangackerGUT} extend the Standard Model (SM) gauge symmetry to larger symmetry groups, provide an explanation for charge quantization and the unification of the three interactions of the SM. 
 GUTs predict baryon number violation and the possibility that protons might decay. 
 In many GUTs, such as $SU(5)$~\cite{GeorgiSU(5), EllisSU(5)} and $SO(10)$~\cite{FritzschSO(10), BabuSO(10)}, the dominant proton decay mode is assumed to be $p\rightarrow e^+\pi^0$.
 
   The theoretical study in ~\cite{Wise3body} showed that three-body proton decays resulting in two $\pi^0$ emitted together with the charged lepton can arise in a general framework with baryon-number–violating interactions, without assuming a specific GUT model.
   In that study, the three-body proton decay into one positron and two pions has a decay rate comparable to that of $p\rightarrow e^+\pi^0$, and similar operators exist for decays with an antimuon in the final state. 
   Another theoretical calculation~\cite{OsetMesonEx} suggests that $p\rightarrow e^+\pi^0\pi^0$ could dominate through the meson exchange mechanism. 
   Motivated by these predictions, a first search for the decay modes $p\rightarrow e^+\pi^0\pi^0$ and $p\rightarrow \mu^+\pi^0\pi^0$ in Super-Kamiokande (SK) data is reported in this paper. These decay modes were previously searched for by the IMB-3 experiment, which reported lower lifetime limits of $1.47\times10^{32}$ years and $1.01\times10^{32}$ years for $p\rightarrow e^+\pi^0\pi^0$ and $p\rightarrow \mu^+\pi^0\pi^0$, respectively~\cite{IMB3}. 
   The analysis presented in this paper is based on techniques used in the previous SK searches for $p\rightarrow e^+\pi^0$ and $p\rightarrow \mu^+\pi^0$ ($p\rightarrow \ell^+\pi^0$)~\cite{Takenaka2020}, which used data from SK-I through SK-IV; here we used data from all of SK's pure water phases, adding the data from SK-V.
     
  This paper is structured as follows: In Section \ref{sec:lab2}, a brief description of the SK detector is given, and the data set used for this analysis is summarized. 
  The simulations of signal and background are summarized in Section \ref{sec:lab3}. 
  The search method and results are discussed in Section \ref{sec:lab4} and Section \ref{sec:lab5}, respectively. In Section \ref{sec:lab6}, the systematic uncertainties of the signal detection efficiencies and the expected number of atmospheric neutrino (atm.-$\nu$) background events are summarized. 
  The calculation of the lower limits on the partial lifetime and the resulting limits are presented in Section \ref{sec:lab7}. Section \ref{sec:lab8} concludes with the search results.

 
\section{\label{sec:lab2}Super-Kamiokande detector and dataset}
  Super-Kamiokande is a 50-kiloton cylindrical water-Cherenkov detector with a height of 41.4 m and a diameter of 39.3 m. It is divided into two regions: the inner detector (ID) and the outer detector (OD). 
  The ID has a diameter of 33.8 m and a height of 36.2 m. It is equipped with 11,129 inward-facing 20-inch photomultiplier tubes (PMTs). 
  The OD is a 2-meter-thick cylindrical shell surrounding the ID and is instrumented with 1,885 outward-facing 8-inch PMTs. The OD serves as a veto for entering cosmic-ray muons and provides shielding from external radiation~\cite{MineTextbook}.

  The SK detector has undergone several upgrades and has operated in various detector configurations, defining the different data-taking phases. 
  Table \ref{tab:tab_SKdet} summarizes the SK detector phases to date~\cite{SK1NIM, SK4NIM, MineTextbook}. 
  Due to an accident during the upgrade work following SK-I, SK-II has only about half of the original number of ID PMTs. 
  To prevent similar accidents, all ID PMTs were encased in fiber-reinforced plastic cases with ultraviolet-transparent acrylic front windows from SK-III onward. 
 For SK-IV, the front-end electronics were upgraded to a system with an ASIC-based high-speed charge-to-time converter (QTC), which became known as the QTC-based electronics with Ethernet (QBEE) module, and was used for both the ID and the OD. 
 The new electronics allowed recording of all PMT signals above a certain signal threshold, which were then sent unfiltered to the readout computers, and events were defined using a software trigger. 
 The QBEE improved the Michel electron tagging efficiency (see Table~\ref{tab:tab_sigeff}) and also enabled neutron tagging for neutrons captured on hydrogen. 
 SK-V was SK's final pure water phase, operated as a dedicated check-up run to verify the detector's performance after detector refurbishments preparing for SK-Gd. 
 To further improve the neutron tagging efficiency, gadolinium sulfate octahydrate was later dissolved in the SK water, marking the start of SK-Gd, which began with SK-VI~\cite{MineTextbook, FirstSKGd, SecondSKGd}. 
 A consistent detector calibration procedure was employed from SK-I to current SK-VIII~\cite{SK4NIM, MineTextbook}. 
  
 For this paper, data from SK-I through SK-V were used.
 The same data processing methods were applied across SK-I to SK-V~\cite{TakeThesis, ThomasThesis}. 
 The total exposure used in this analysis is 0.401 megaton-years.

\begin{table*}[ht]

\centering
\footnotesize
\setlength{\tabcolsep}{4pt}
\caption{\footnotesize Summary of the SK detector phases~\cite{SK1NIM, SK4NIM, MineTextbook}. This analysis uses data from the pure-water phases called SK-I to SK-V. ``FC'', ``FV'', and ``ATM'' stand for fully contained, fiducial volume, and analog timing module, respectively. FC, FV used in this analysis will be explained in Section~\ref{sec:lab4}.}

\label{tab:tab_SKdet}
\resizebox{0.9\textwidth}{!}{

\begin{tabular}{lcccccc}
\toprule
\textbf{Phase} & 
\textbf{Dates} &
\makecell{\textbf{FC Live Time} \\ \textbf{[days]}} & 
\makecell{\textbf{ID PMT} \\ \textbf{Photo-Coverage [\%]}} & 
\textbf{ID PMT Cover} & 
\makecell{\textbf{ID Front-End} \\ \textbf{Electronics Module}} & 
\makecell{\textbf{Gd-Loading : } \\ \textbf{n-capture on Gd [\%]}} \\
\midrule
I  & 1996--2001 & 1489.19 & 40 & No  & ATM  & No \\
II & 2002--2005 & 798.59  &  19 & Yes & ATM  & No \\
III & 2006--2008 & 518.08  & 40 & Yes & ATM  & No \\
IV & 2008--2018 & 3244.4 & 40 & Yes & QBEE & No \\
V & 2019--2020 & 461.0  & 40 & Yes & QBEE & No \\
VI & 2020--2020 & 564.4  & 40 & Yes & QBEE & Yes ($\sim$50) \\
VII & 2022--2023 & 405.4  & 40 & Yes & QBEE & Yes ($\sim$75) \\
VIII & 2023--Present & Ongoing   & 40 & Yes & QBEE & Yes ($\sim$75) \\
\bottomrule
\end{tabular}
}
\end{table*}

\section{\label{sec:lab3}Simulation}

 For each detector phase, dedicated Monte Carlo (MC) samples for the two decay modes, $p \rightarrow e^+\pi^0\pi^0$ and $p \rightarrow \mu^+\pi^0\pi^0$, were generated as well as atm.-$\nu$ background samples. 
 From these samples, the signal detection efficiency and the expected number of background events were estimated independently for each of the five detector phases. 
 
 For signal MC samples, the initial three-body decay kinematics were simulated as well as the pion's final state interactions ($\pi$-FSI) within the oxygen nuclei in the water. 
 Free protons, the hydrogen nuclei in water, do not interact with other nucleons, whereas bound protons in the oxygen nucleus are subject to the effects of Fermi motion~\cite{Fermimotion}, nuclear binding energy~\cite{BindingE}, and correlations with other nucleons (correlated decay)~\cite{Corrdecay}. 
 The $\pi$-FSI—absorption, scattering, and charge exchange—were simulated using the NEUT package~\cite{NEUTHayato, NEUTMitsuka, AIP2008}. 
 These physics processes contributed to the systematic uncertainties of the signal detection efficiencies for each data-taking period, as discussed in Section~\ref{sec:lab6}.

\begin{figure}[hbp]
\centering
\begin{tabular}{cc}
\includegraphics[width=4.0cm]{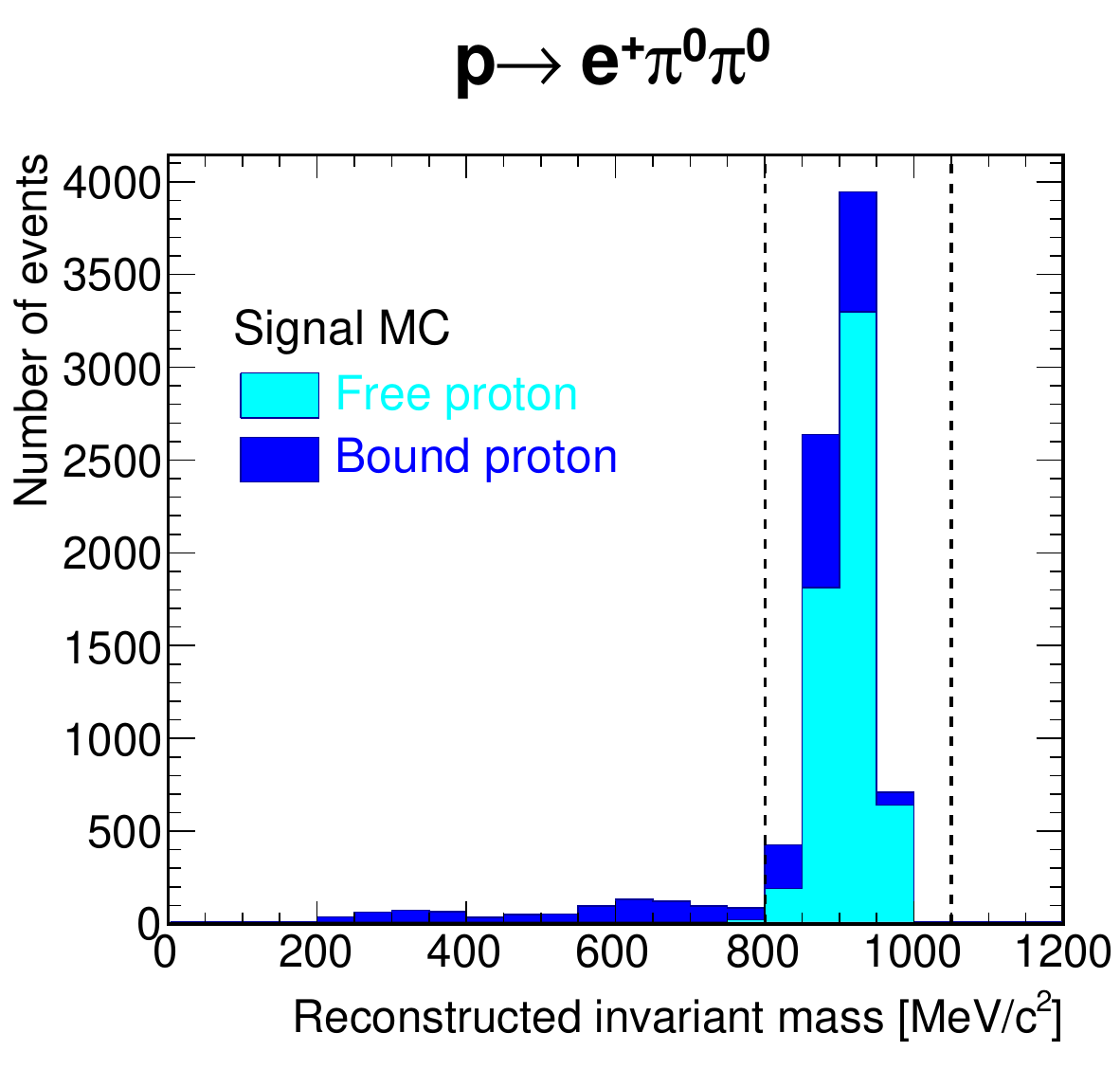} &
\includegraphics[width=4.0cm]{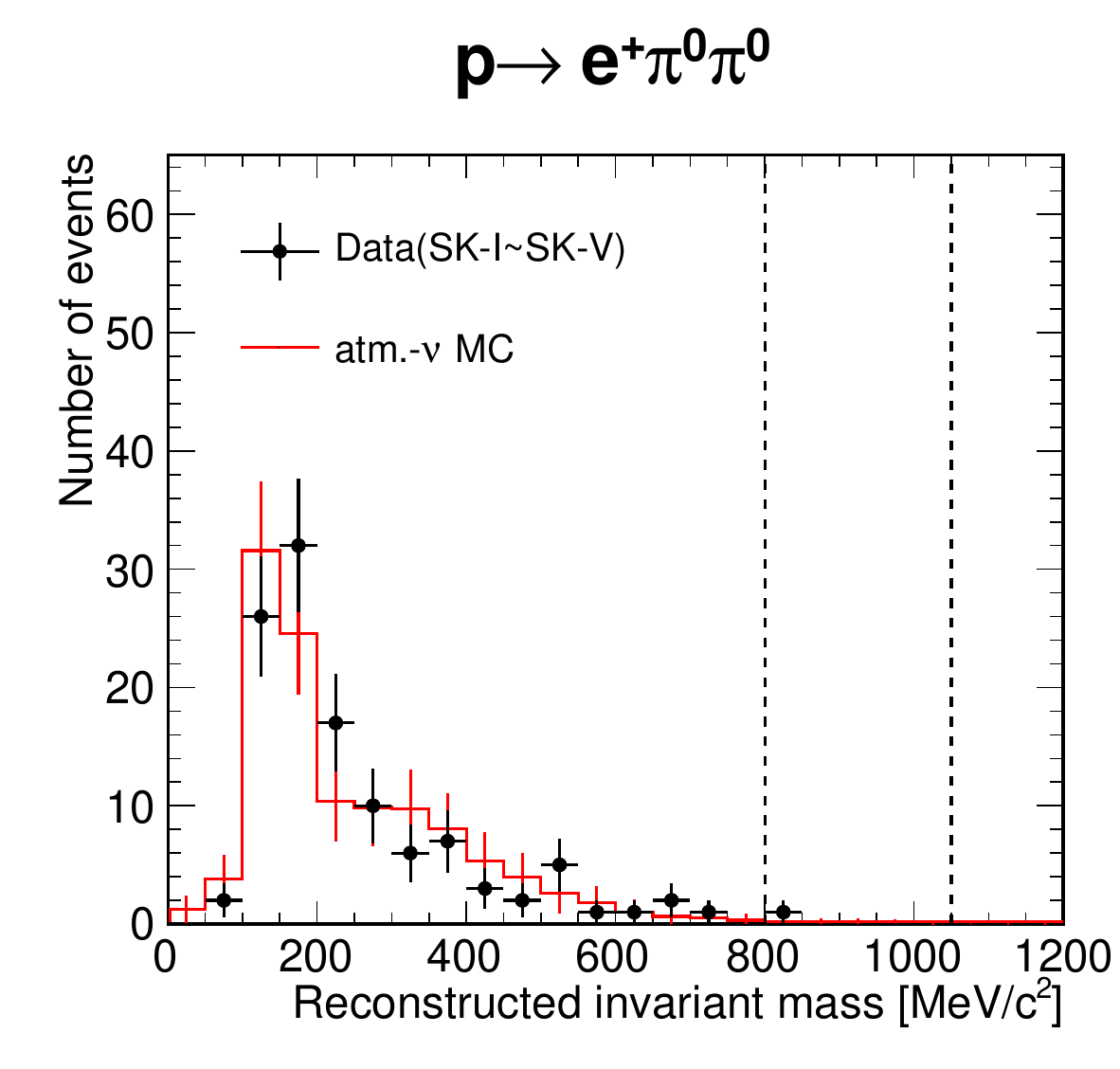} \\
\includegraphics[width=4.0cm]{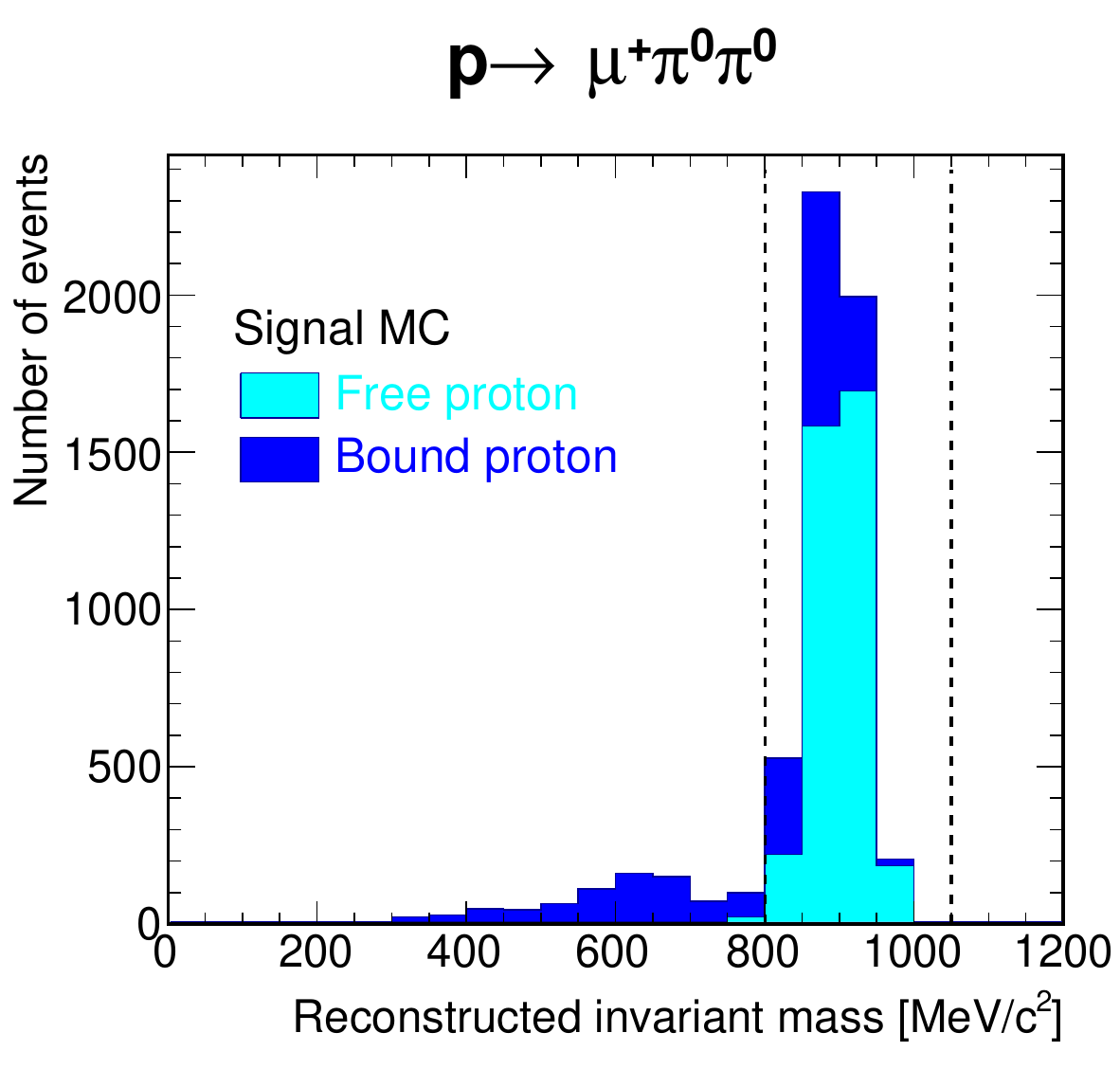} &
\includegraphics[width=4.0cm]{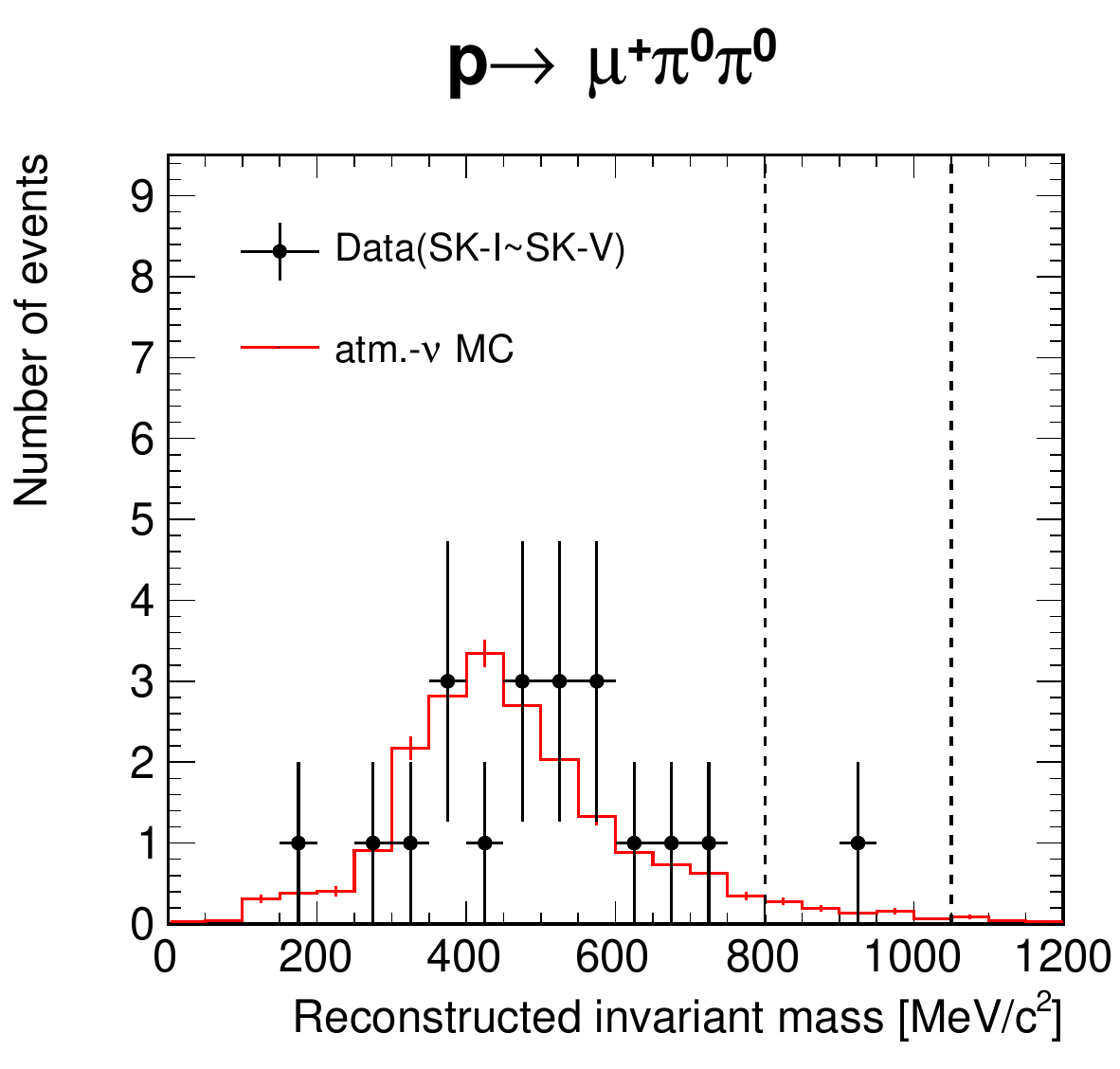} \\
\end{tabular}
\caption{\footnotesize The reconstructed invariant mass distributions after applying all the event selection criteria except for the reconstructed invariant mass cut, for the signal MC (left) and the atm.-$\nu$ background MC and the data (right) in SK-I to SK-V, for $p\rightarrow e^+\pi^0\pi^0$ (upper) and $p\rightarrow \mu^+\pi^0\pi^0$ (bottom). The atm.-$\nu$ background MC (red histogram) is scaled to the total livetime for each SK phase. In the signal MC distribution, free protons are shown in cyan, and bound protons are shown in blue histograms. The signal region is defined between the two vertical dashed lines.}
\label{fig:fig_totM}
\end{figure}

\begin{figure}[hbp]
\centering
\begin{tabular}{cc}
\includegraphics[width=4.0cm]{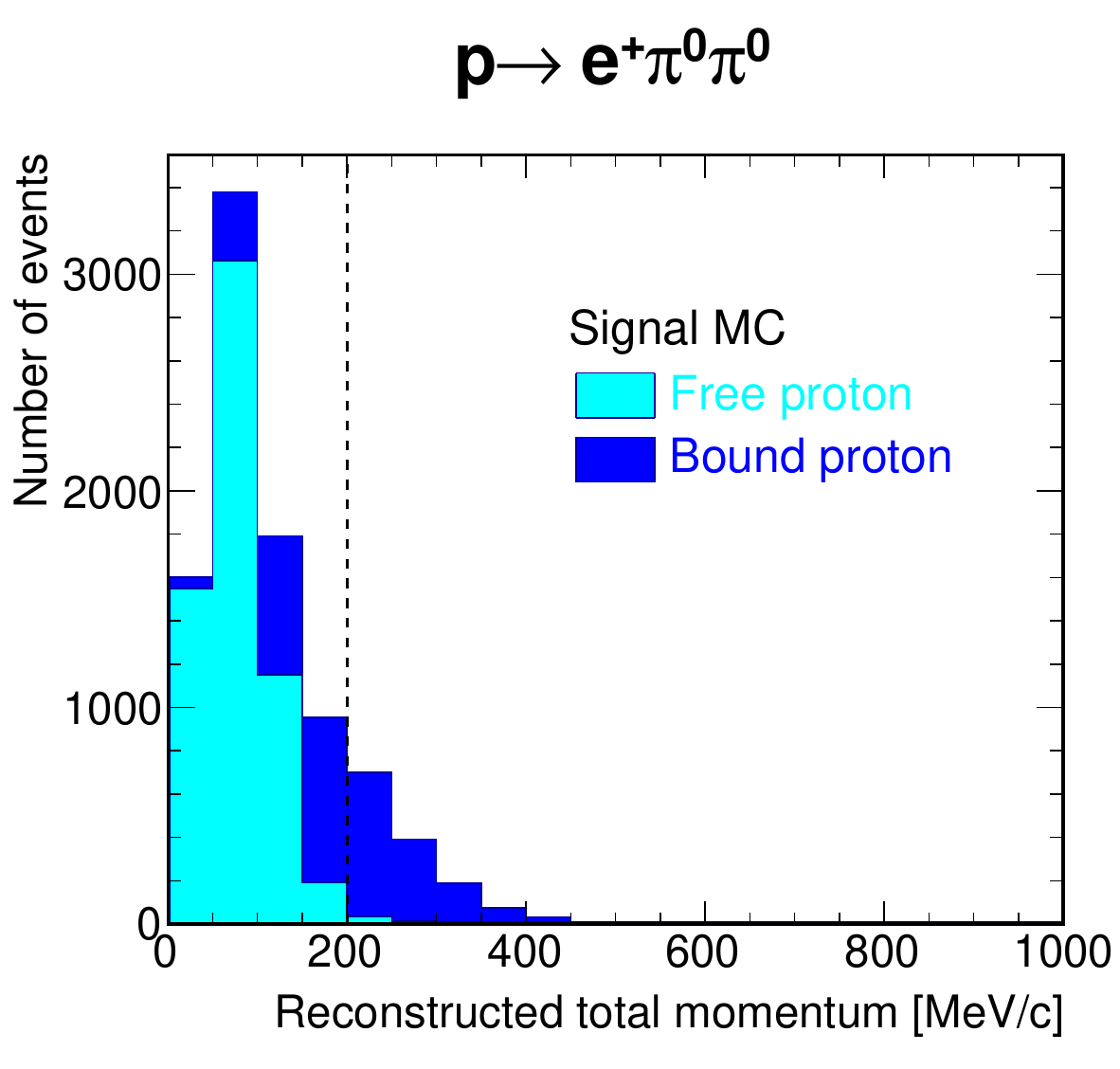} &
\includegraphics[width=4.0cm]{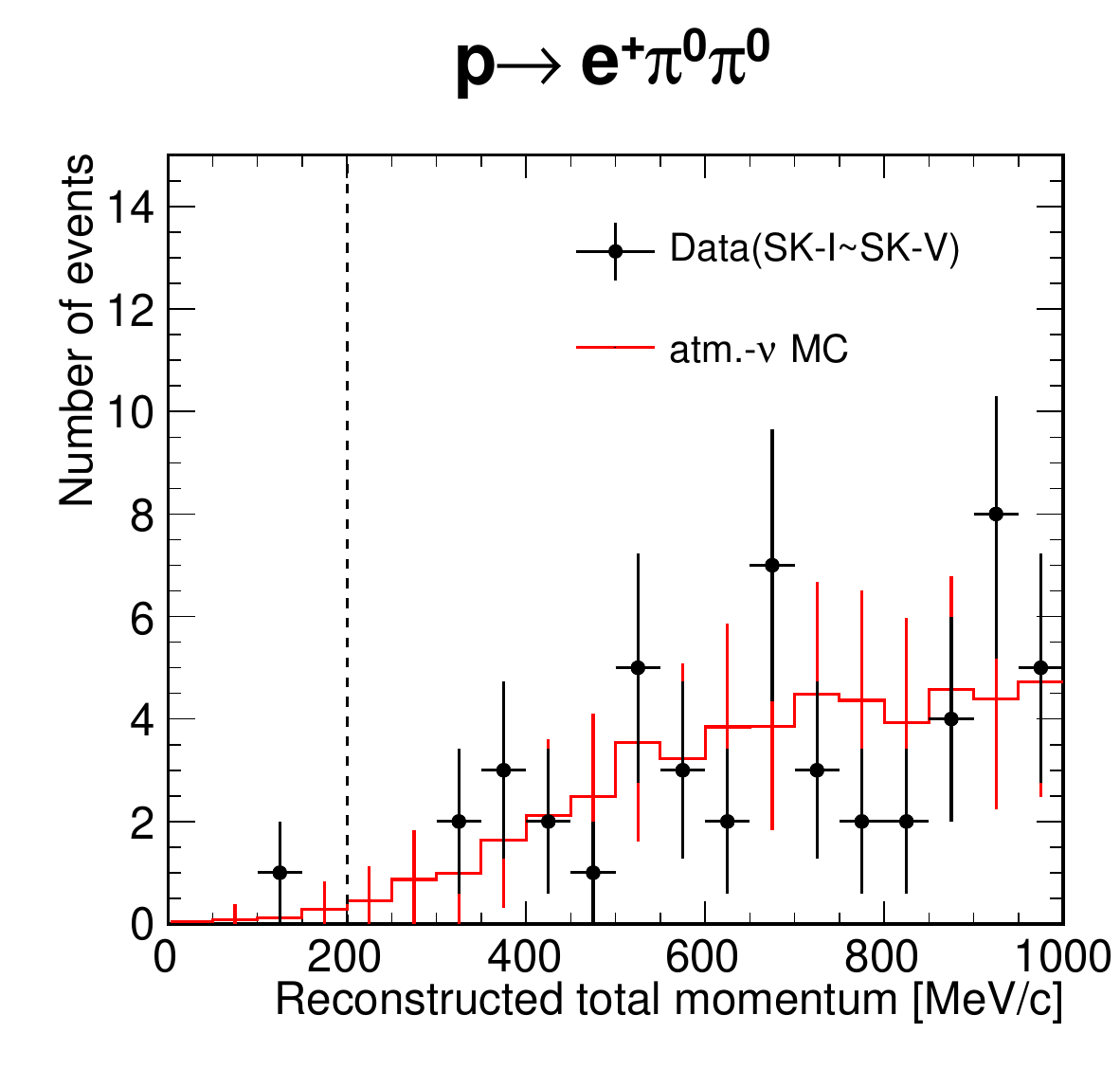} \\
\includegraphics[width=4.0cm]{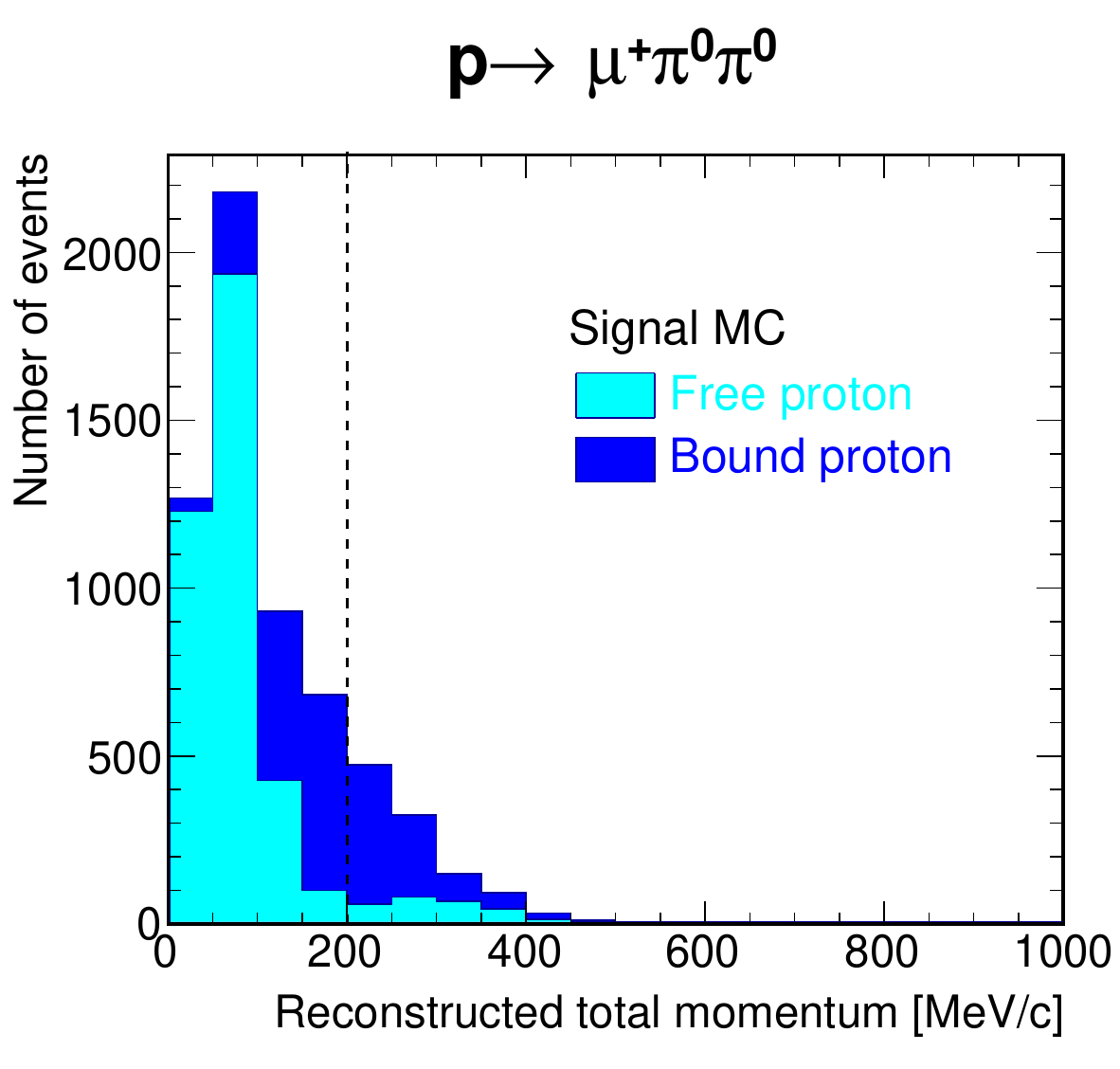} &
\includegraphics[width=4.0cm]{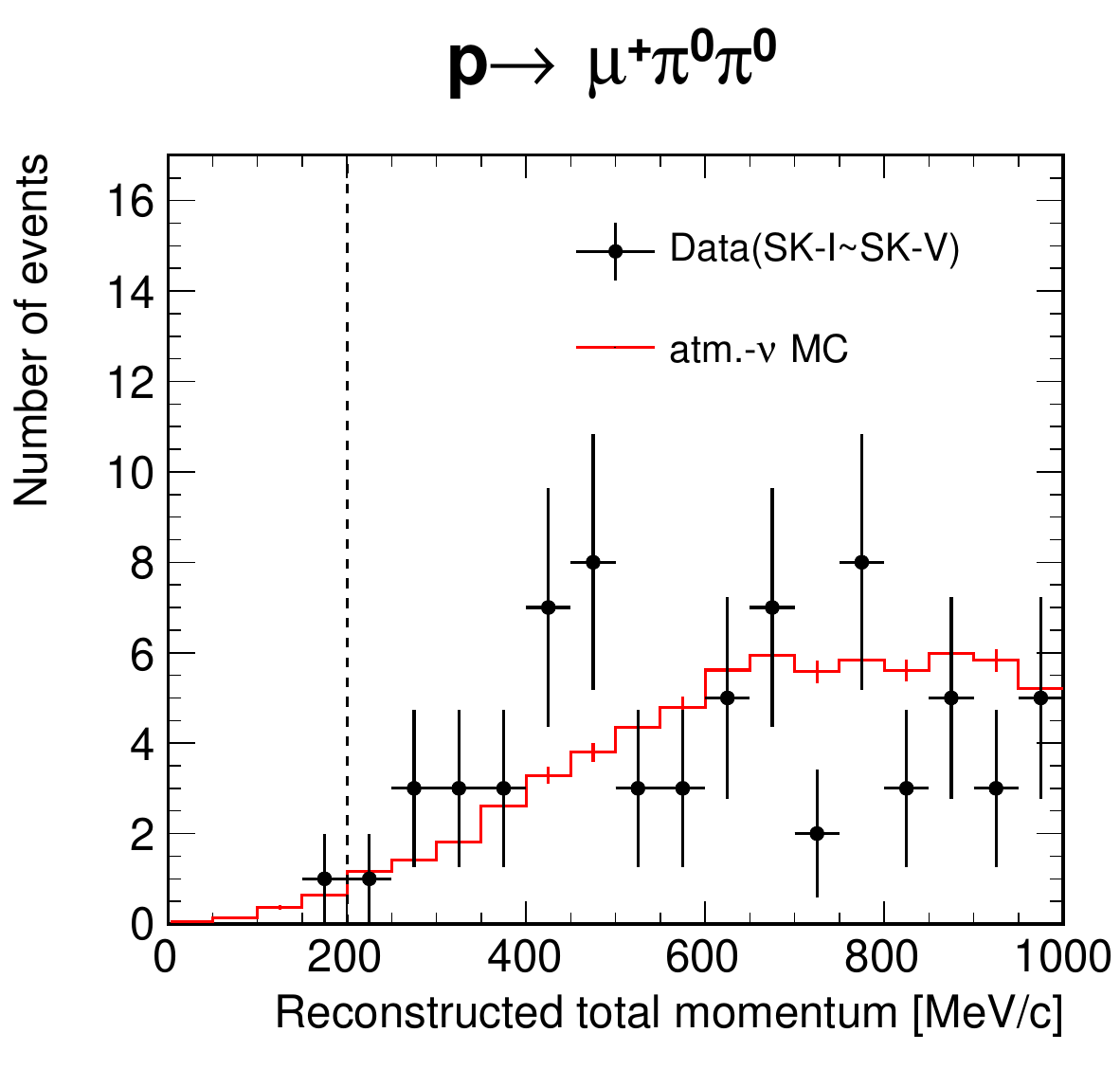} \\
\end{tabular}
\caption{\footnotesize The total momentum distribution after applying all event selection criteria except for the total momentum cut. The left panel shows the signal MC and the right panel shows the atm.-$\nu$ background MC (red histogram) and the data in SK-I to SK-V (black crosses) for $p\rightarrow e^+\pi^0\pi^0$ (upper) and $p\rightarrow \mu^+\pi^0\pi^0$ (bottom). In the signal MC distribution, the cyan histogram represents free protons and the blue histogram represents bound protons. The atm.-$\nu$ background MC (red histogram) is scaled to the total livetime for each SK phase. The signal region is defined as to the left of the vertical dashed line.}
\label{fig:fig_totP}
\end{figure}

 Background events were generated by atm.-$\nu$ interactions in the inner detector, and events entering the analysis were selected within the fiducial volume (FV).
 For the atm.-$\nu$ background MC, the Honda flux model~\cite{Hondaflux} was used, and neutrino interactions on oxygen nuclei and free protons in water were simulated using NEUT, with detector responses simulated by the GEANT3-based SK detector simulation~\cite{GEANT3}.
  As for the signal, those physics processes contribute to systematic uncertainties of the expected number of atm.-$\nu$ background events. 500 years of atm.-$\nu$ background MC events were generated for each SK phase, and scaled to the livetime.
 
  The detector simulation calculated the amount of light detected at each PMT and its timing from the Cherenkov light generation by charged particles in water, followed by light propagation in water and reflection at the PMTs, their housing, and the detector's structural surfaces. 
  To this simulated signal, PMT dark noise was added and the resulting waveforms were processed through a model of the electronics' response to complete the detector simulation. 
 These processes were treated as the sources of detector-related uncertainties in the estimation of the systematic uncertainties of the signal detection efficiency and the expected number of atm.-$\nu$ background events. 
 All detector-related uncertainties were evaluated separately for each SK phase using phase-specific calibration data, while the definitions of the uncertainty sources and the estimation procedures were common to all SK phases. 
 The detector simulation used in this analysis follows that of Ref.~\cite{Takenaka2020}, with updates implemented for each detector phase.

\begin{table*}[htbp]
\centering
\caption{\footnotesize Signal detection efficiency for $p\rightarrow e^+\pi^0\pi^0$ and $p\rightarrow \mu^+\pi^0\pi^0$ for event selections per SK detector phase. The errors are the MC statistical errors only. These selection criteria are defined in Section~\ref{sec:lab4}.}
\label{tab:tab_sigeff} 
\renewcommand{\arraystretch}{1.2}
\footnotesize
\resizebox{0.8\textwidth}{!}{
\begin{tabular}{c|ccccc|ccccc}
\toprule
\multicolumn{1}{c|}{} & \multicolumn{5}{c|}{$p\rightarrow e^+\pi^0\pi^0$ Signal detection efficiency [\%]} & \multicolumn{5}{c}{$p\rightarrow \mu^+\pi^0\pi^0$ Signal detection efficiency [\%]} \\
\cmidrule(lr){2-6} \cmidrule(lr){7-11}
Cuts & SK-I & SK-II & SK-III & SK-IV & SK-V & SK-I & SK-II & SK-III & SK-IV & SK-V \\
\midrule
C1 & 99.5 $\pm$ 1.1 & 99.4 $\pm$ 1.1 & 96.5 $\pm$ 1.1 & 96.8 $\pm$ 1.1 & 97.8 $\pm$ 1.1 & 95.3 $\pm$ 1.5 & 96.0 $\pm$ 1.5 & 93.2 $\pm$ 1.5 & 94.5 $\pm$ 1.5 & 95.2 $\pm$ 1.5 \\
C2 & 58.3 $\pm$ 0.8 & 58.8 $\pm$ 0.8 & 58.1 $\pm$ 0.8 & 58.7 $\pm$ 0.8 & 59.8 $\pm$ 0.8 & 55.1 $\pm$ 1.0 & 54.7 $\pm$ 1.0 & 55.5 $\pm$ 1.0 & 56.0 $\pm$ 1.0 & 57.4 $\pm$ 1.1 \\
C3 & 45.7 $\pm$ 0.7 & 40.6 $\pm$ 0.7 & 46.5 $\pm$ 0.8 & 46.5 $\pm$ 0.8 & 46.7 $\pm$ 0.8 & 37.4 $\pm$ 0.8 & 34.3 $\pm$ 0.8 & 37.9 $\pm$ 0.8 & 37.9 $\pm$ 0.8 & 39.1 $\pm$ 0.8 \\
C4 & 44.5 $\pm$ 0.7 & 39.3 $\pm$ 0.7 & 45.0 $\pm$ 0.7 & 44.6 $\pm$ 0.7 & 44.9 $\pm$ 0.7 & 29.4 $\pm$ 0.7 & 26.6 $\pm$ 0.6 & 29.8 $\pm$ 0.7 & 34.9 $\pm$ 0.8 & 36.0 $\pm$ 0.8 \\
C5 & 44.5 $\pm$ 0.7 & 39.3 $\pm$ 0.7 & 45.0 $\pm$ 0.7 & 43.6 $\pm$ 0.7 & 44.2 $\pm$ 0.7 & 29.4 $\pm$ 0.7 & 26.6 $\pm$ 0.6 & 29.8 $\pm$ 0.7 & 34.2 $\pm$ 0.7 & 35.3 $\pm$ 0.8 \\
C6 & 22.8 $\pm$ 0.5 & 19.4 $\pm$ 0.5 & 23.4 $\pm$ 0.5 & 21.5 $\pm$ 0.5 & 22.4 $\pm$ 0.5 & 14.3 $\pm$ 0.4 & 12.3 $\pm$ 0.4 & 14.0 $\pm$ 0.4 & 16.8 $\pm$ 0.5 & 16.9 $\pm$ 0.5 \\
C7 & 19.2 $\pm$ 0.5 & 16.4 $\pm$ 0.4 & 19.8 $\pm$ 0.5 & 18.3 $\pm$ 0.5 & 19.1 $\pm$ 0.5 & 11.7 $\pm$ 0.4 & 9.8 $\pm$ 0.4 & 11.7 $\pm$ 0.4 & 13.9 $\pm$ 0.4 & 14.1 $\pm$ 0.4 \\
\bottomrule
\end{tabular}}
\end{table*}

\begin{figure*}[hbtp]
\begin{tabular}{cc}
\includegraphics[width=18cm]{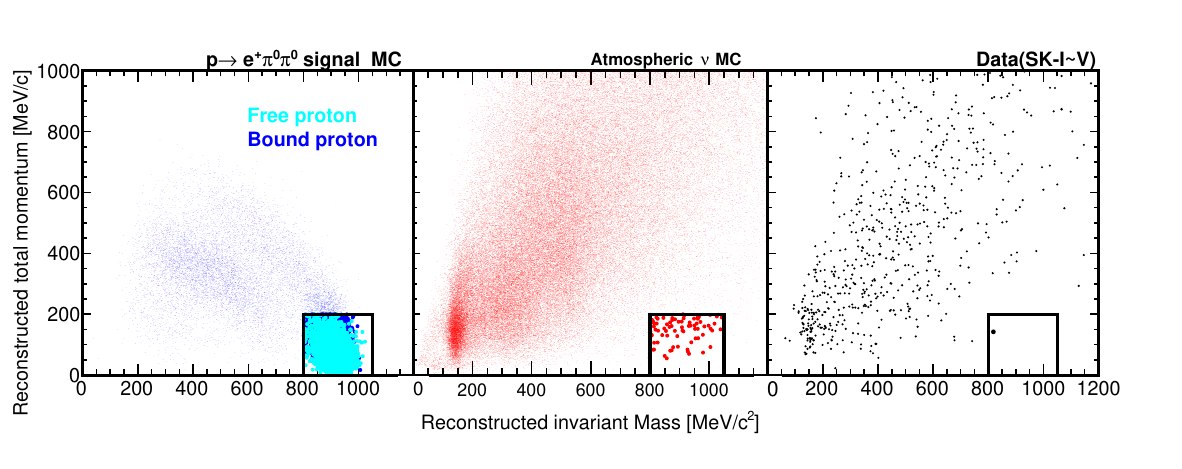} \\
\includegraphics[width=18cm]{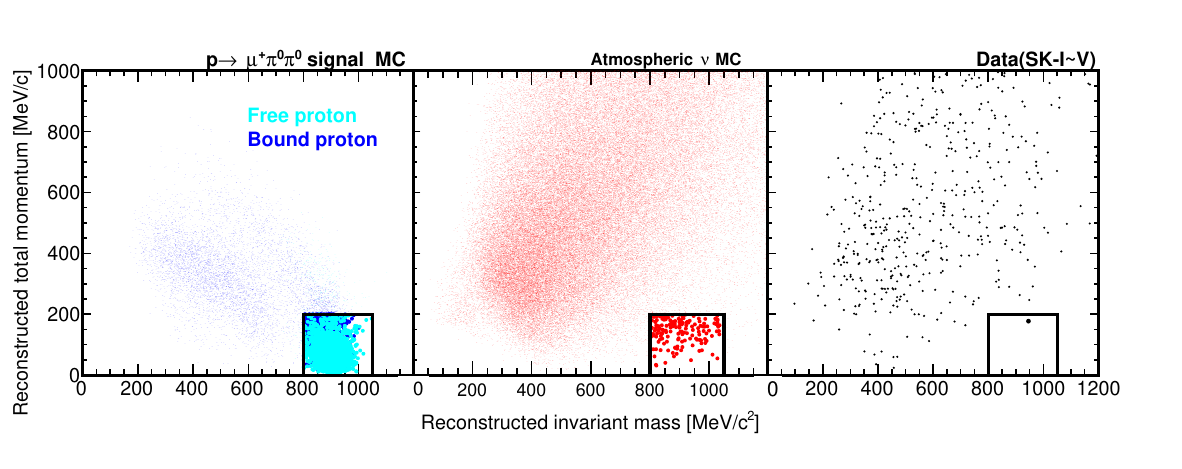} &
\end{tabular}
\caption{\footnotesize Reconstructed invariant masses versus total momentum distribution for $p\rightarrow e^+\pi^0\pi^0$ (upper) and $p\rightarrow \mu^+\pi^0\pi^0$ (bottom) after applying all selection criteria except for the reconstructed invariant mass and total momentum cuts in SK-I to SK-V. The left panel shows the signal MC, the center panel shows the atm.-$\nu$ background MC, and the right panel shows the data distributions. The black box in each panel represents the signal region and the markers in the signal region have been enlarged for visibility. In the signal MC, the cyan dots represent free protons, and the blue dots represent bound protons. }
\label{fig:fig_scatter}
\end{figure*}

\begin{figure}[hbt]
\centering
\begin{tabular}{cc}
\includegraphics[width=4.0cm]{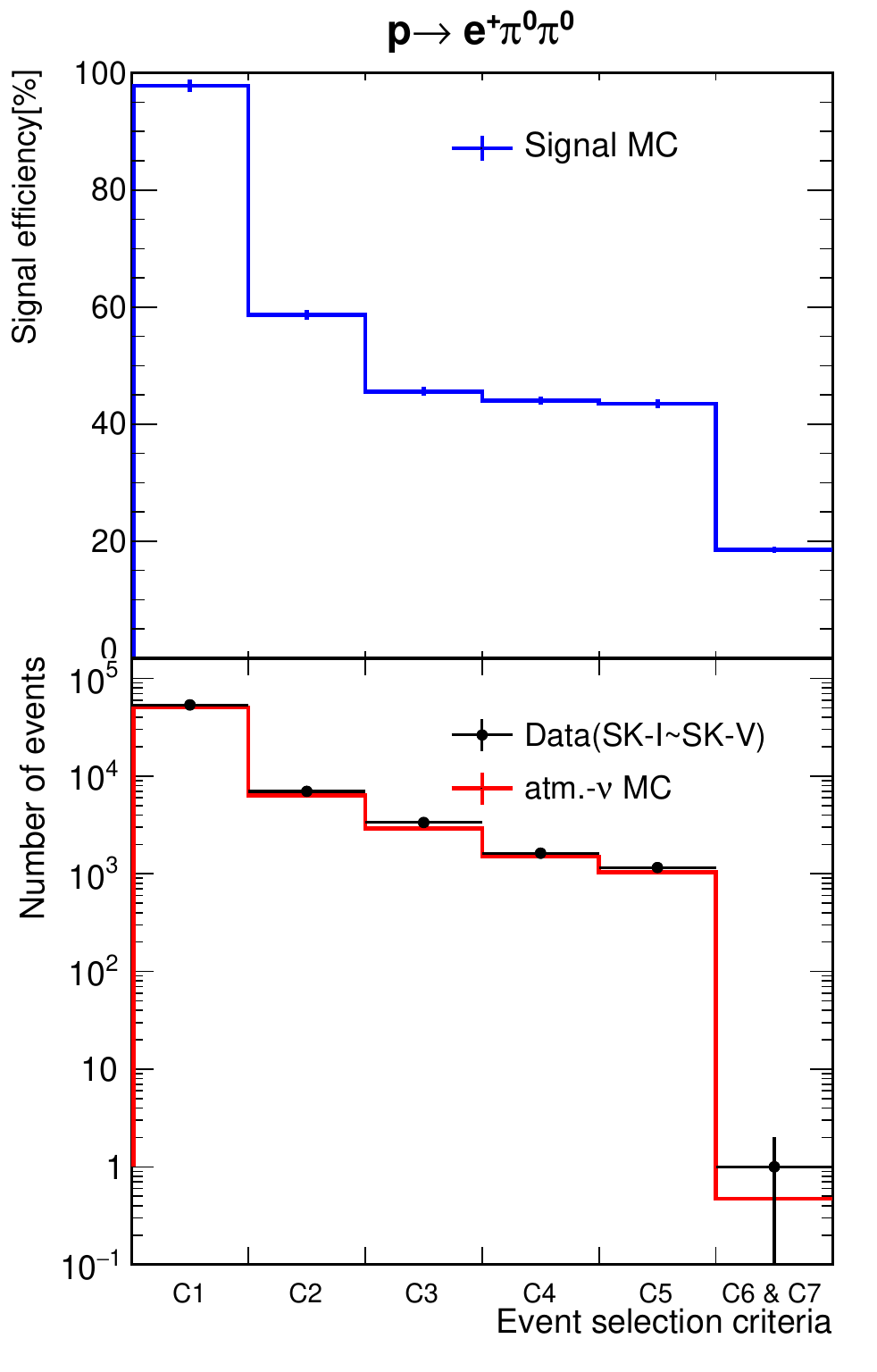} &
\includegraphics[width=4.0cm]{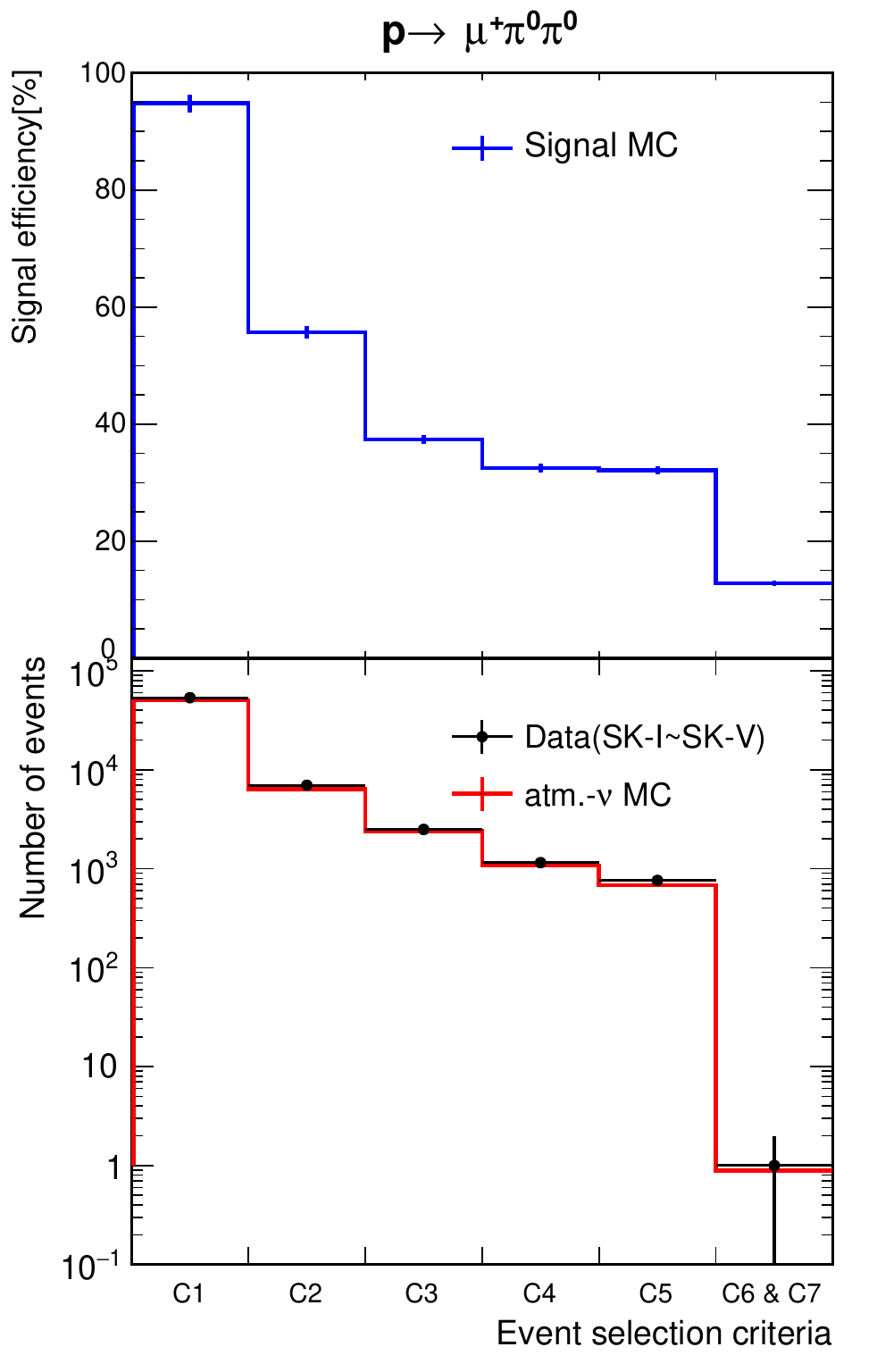} 
\end{tabular}
\caption{\footnotesize The event reductions along the event selections for signal MC (blue), the atm.-$\nu$ background MC (red), and data (black points) for $p\rightarrow e^+\pi^0\pi^0$  (left) and $p\rightarrow \mu^+\pi^0\pi^0$ (right). Shown are the signal efficiency, weighted by the livetime, and the total number of background events from SK-I to SK-V. For the background, the MC for each SK phase was scaled to that phase's livetime. The event selection criteria are defined in Section~\ref{sec:lab4}. The statistical errors on each bin are shown. }
\label{fig:fig_evtreduction}
\end{figure}

\begin{figure*}[ht]
\includegraphics[width=7cm]{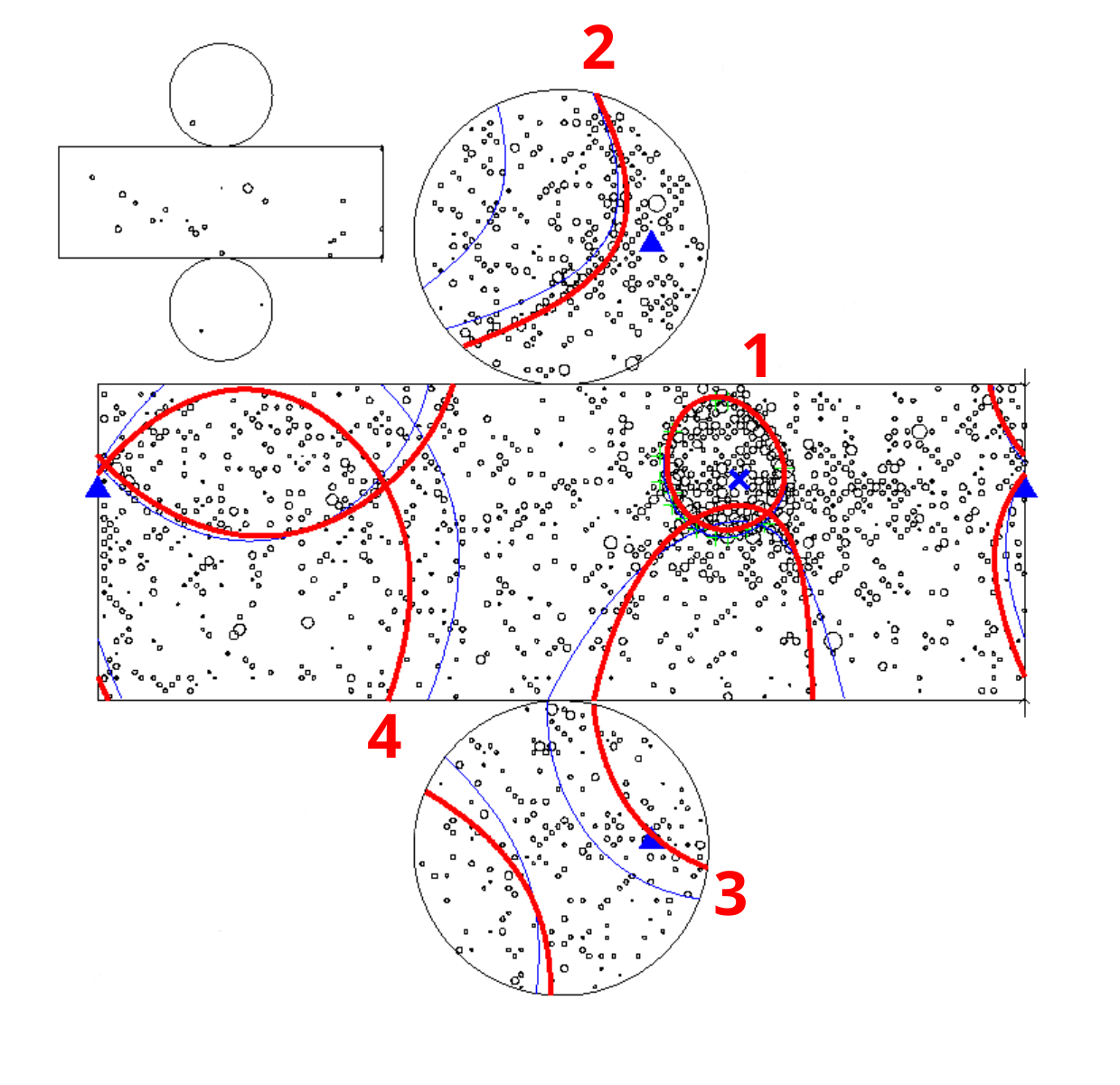} 
\includegraphics[width=7cm]{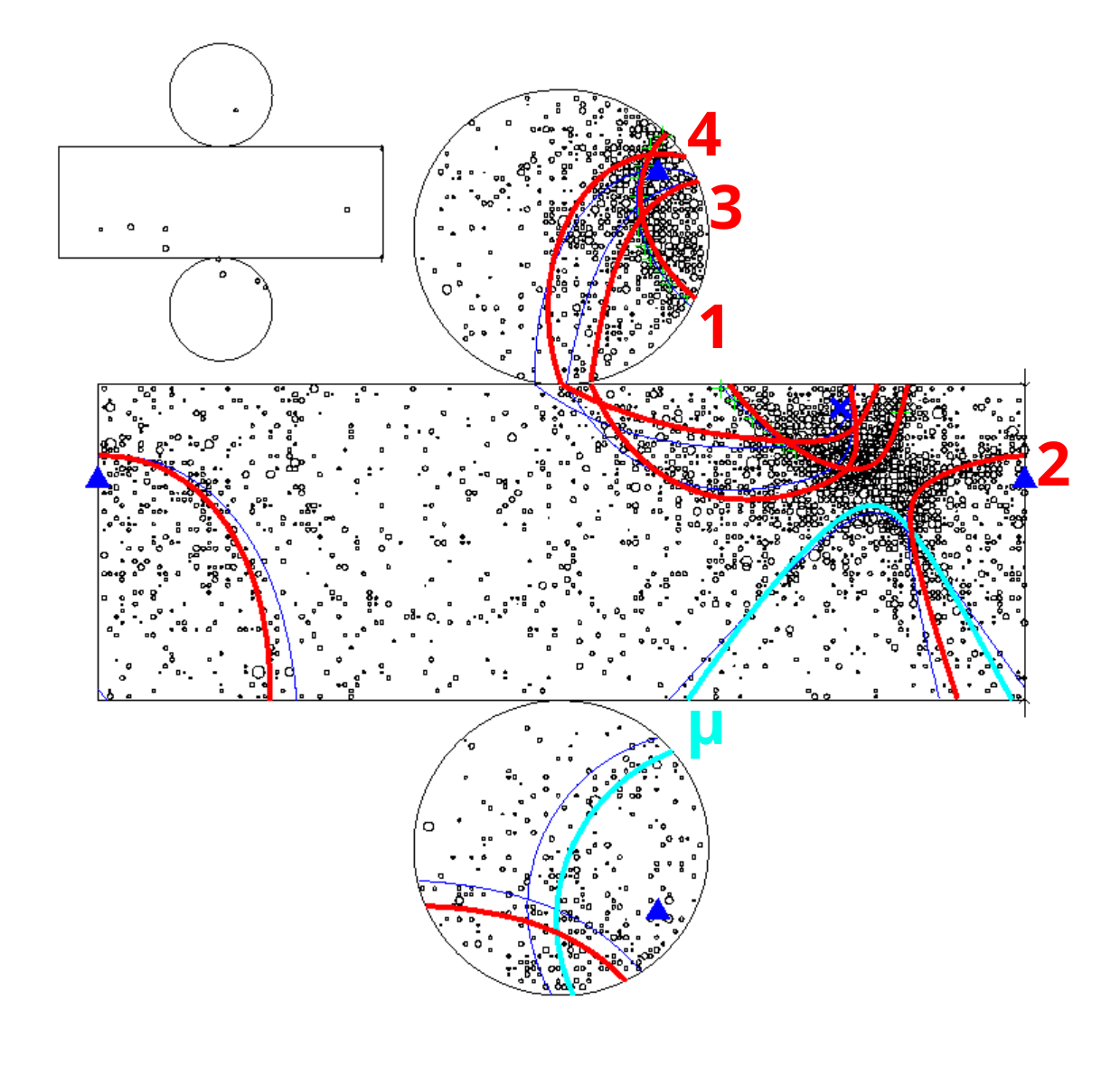} 
\caption{\footnotesize Event displays for data candidates of $p\rightarrow e^+\pi^0\pi^0$ in SK-II (left) and $p\rightarrow \mu^+\pi^0\pi^0$  in SK-IV (right). These show expanded views of the vertical cylindrical SK tanks, with the ID (center) and OD (left-top). Each open circle represents a hit PMT, with its diameter proportional to the detected photoelectrons. The blue filled triangles in ID mark the reconstructed vertex. The solid rings indicate reconstructed ring edges. The red reconstructed rings represent showering particle ($e$-like), while the cyan reconstructed ring represents a non-showering particle ($\mu$-like). The numbering of the red rings corresponds to the reconstructed rings described in Section~\ref{sec:lab5}.}
\label{fig:fig_evtdisp}
\end{figure*}

\section{\label{sec:lab4}Search methods}

 This analysis used events termed “fully contained” (FC), in which all visible particles are contained within the ID, the reconstructed vertex lies in the fiducial volume (FV), and no cluster of hits is observed in the OD. 
The FV was defined as the part of the inner detector that lies at least 2\,m from the nearest ID wall, containing a total target mass of 22.5~kilotons (approximately $7.5\times10^{33}$ protons).
 The FC and FV selections provide the suppression of low-energy background from radioactive decays and cosmic-ray muons using multi-step data reduction methods developed for the atm.-$\nu$ oscillation and proton decay analyses~\cite{TakeThesis, ThomasThesis}. 
 After these FC and FV selections, events are required to have visible energy above 30 MeV. 

An event reconstruction algorithm called APFit~\cite{SK1NIM, MineTextbook, Shiozawa1999} estimated kinematic properties such as the vertex position, the number of Cherenkov rings, the ring directions and momenta, and particle types. 
Showering particles (typically electrons and gammas) produce diffused ring patterns due to the electromagnetic showers, while non-showering particles (such as muons) yield rings with clear edges. 
The particle identification algorithm is based on PMT hit patterns and Cherenkov cone opening angles. 
Reconstructed total momentum and invariant mass are calculated using the reconstructed momenta and energies of the final-state particles. 
In addition, Michel electrons from $\mu$ decays are tagged, and neutrons are identified via the 2.2~MeV $\gamma$ from neutron capture on hydrogen (available in SK-IV and later); these tags are used in the event selection to suppress atm.-$\nu$ backgrounds.
These variables are then used in evaluating the following event selection criteria:

 \medskip
 
 \begin{enumerate}[
    label=C\arabic*.,      
    leftmargin=*,          
    itemsep=0.3\baselineskip,  
    topsep=0pt,            
    parsep=0pt
]
  \item Events must be fully contained with their reconstructed vertex
        within the fiducial volume.
  \item Events must have three, four, or five reconstructed
        Cherenkov rings.
  \item All of the event's rings must be reconstructed as showering for
        $p\rightarrow e^+\pi^0\pi^0$ and one of the event's rings must be
        non-showering for $p\rightarrow \mu^+\pi^0\pi^0$.
  \item There must be no tagged Michel electrons associated with the event for
        $p\rightarrow e^+\pi^0\pi^0$ and one for
        $p\rightarrow \mu^+\pi^0\pi^0$.
  \item For SK-IV and SK-V data, there must be no tagged neutrons associated with the event.
  \item The event's reconstructed invariant mass must be between
        800 and 1050 MeV/$c^{2}$.
  \item The event's reconstructed total momentum must be less than
        200 MeV/$c$.
\end{enumerate}
\medskip

 Since the particles ($e$, $\mu$, and $\pi^0$) in the final state are the same as in the latest $p\rightarrow \ell^+\pi^0$ analysis~\cite{Takenaka2020}, the same cut parameters and similar event selections were applied in this analysis. 
 Whereas the analysis in Ref.~\cite{Takenaka2020} used an expanded FV (with the minimum distance from the reconstructed vertex to the nearest ID wall greater than 1 m), this analysis used the conventional FV, where the distance is greater than 2 m. 
  In the additional FV (the distance between 1 m and 2 m from the wall), the reconstructed vertex is located closer to the detector boundary, which increases the probability that low-energy gammas are not reconstructed or are reconstructed as a single ring due to overlapping, thereby reducing the reconstructed ring multiplicity. 
  In particular, for free-proton decays, the fraction of signal events rejected by the ring-counting cut (C2) is less than 1\% in the conventional FV, whereas it rises to approximately 20\% in the additional FV. Consequently, only the conventional FV is used in this analysis.
 There is no cut applied to the reconstructed $\pi^0$ mass since we accept events where some of the gammas are undetected, and the fraction of mis-combinations (incorrect pairing of two reconstructed gamma candidates to form a $\pi^0$) from the two $\pi^0$ decays was found to be about 40\%.
 In the latest $p\rightarrow \ell^+\pi^0$ analysis~\cite{Takenaka2020}, a two-box method was used to obtain better sensitivity; it separates events into low- and high-momentum regions based on the total momentum. In this analysis, about 20\% of the free proton decay events exceeded the two-box method threshold (100 MeV/$c$) as shown in Figure~\ref{fig:fig_totP}, making it difficult to separate the free proton events. Therefore, a single-box method (selecting events by a single cut threshold on the total momentum) was used. 
 The single-box upper cut (an upper limit of 200 MeV/$c$ on the reconstructed total momentum) was optimized to maximize the sensitivity in this analysis.

\begin{table}[hbp]
\caption{\footnotesize Breakdown of the neutrino interaction modes of the atm.-$\nu$ background events remaining in the signal regions from SK-I to SK-V. The respective neutrino interactions are listed in the first column. CC and NC denote charged-current and neutral-current. QE, $1\pi$, multi-$\pi$, $\eta$, $K$, and DIS correspond to quasi-elastic scattering, single $\pi$ production, multi-$\pi$ production, single $\eta$ production, single kaon production, and deep inelastic scattering, respectively. The second and third columns show the fractions (with statistical errors) of atm.-$\nu$ background events for each neutrino interaction mode for $p\rightarrow e^+\pi^0\pi^0$ and $p\rightarrow \mu^+\pi^0\pi^0$, respectively.}
\label{tab:tab_bkgint}
\resizebox{0.8\columnwidth}{!}{
\centering
\begin{tabular}{cccc}
\toprule
\multicolumn{2}{c}{Neutrino interaction} & $p\rightarrow e^+\pi^0\pi^0$ [\%] & $p\rightarrow \mu^+\pi^0\pi^0$ [\%] \\
\midrule
\multicolumn{2}{c}{CCQE} & 7 $\pm$ 4 & 6 $\pm$ 2 \\
\multirow{2}{1.5cm}{CC single} & $\pi^+$ & 16 $\pm$ 4 & 23 $\pm$ 5 \\
 & $\pi^0$ & 14 $\pm$ 6 & 7 $\pm$ 3 \\
\multicolumn{2}{c}{CC multi-$\pi$} & 22 $\pm$ 7 & 29 $\pm$ 5 \\
\multicolumn{2}{c}{CC $\eta$} & 5 $\pm$ 2 & 2 $\pm$ 2 \\
\multicolumn{2}{c}{CC $K$} & 0 & 2 $\pm$ 2 \\
\multicolumn{2}{c}{CC DIS} & 0 & 9 $\pm$ 3 \\
\multirow{2}{1.5cm}{NC single} & $\pi^0$ & 1 $\pm$ 1 & 1 $\pm$ 1 \\
 & $\pi^-$ & 1 $\pm$ 1 & 1 $\pm$ 1 \\
\multicolumn{2}{c}{NC multi-$\pi$} & 15 $\pm$ 5 & 8 $\pm$ 3 \\
\multicolumn{2}{c}{NC $K$} & 2 $\pm$ 6 & 0 $\pm$ 1 \\
\multicolumn{2}{c}{NC DIS} & 17 $\pm$ 6 & 12 $\pm$ 4 \\
\multicolumn{2}{c}{NCQE} & 0 & 0 \\
\midrule
\multicolumn{2}{c}{Total} & 100 & 100 \\
\bottomrule
\end{tabular}
}
\end{table}

\begin{table*}[hbtp]
\centering
\caption{\footnotesize Summary of the search for the proton decays, $p \rightarrow e^+\pi^0\pi^0$ and $p \rightarrow \mu^+\pi^0\pi^0$. The error shown for each contribution is the statistical error.}
\label{tab:tab_summary}
\resizebox{0.9\textwidth}{!}{
\begin{tabular}{l|ccccc||ccccc}
\toprule
& \multicolumn{5}{c||}{$p \rightarrow e^+\pi^0\pi^0$} & \multicolumn{5}{c}{$p \rightarrow \mu^+\pi^0\pi^0$} \\
\cmidrule(lr){2-6} \cmidrule(lr){7-11}
 & SK-I & SK-II & SK-III & SK-IV & SK-V & SK-I & SK-II & SK-III & SK-IV & SK-V  \\
\midrule
Exposure [Mton $\cdot$ year] & 0.092 & 0.049 & 0.032 & 0.200 & 0.028  & 0.092 & 0.049 & 0.032 & 0.200 & 0.028  \\
Signal efficiency [\%] & $19.2 \pm 0.5$  & $16.4 \pm 0.4$ & $19.8 \pm 0.5$ & $18.3 \pm 0.5$ & $19.1 \pm 0.5$   & $11.7 \pm 0.4$ & $9.8 \pm 0.4$ & $11.7 \pm 0.4$  & $13.9 \pm 0.4$  & $14.1 \pm 0.4$ \\
Signal syst. uncertainty [\%] & $9.9 \pm 0.8$ & $12.2 \pm 1.0$ & $13.5 \pm 0.9$ & $8.9 \pm 0.5$ & $13.1 \pm 0.9$   & $12.3 \pm 0.4$ & $13.2 \pm 0.5$ & $11.2 \pm 0.5$ & $12.4 \pm 0.4$ & $12.7 \pm 0.4$  \\
Background rate [events] & $0.17 \pm 0.04$ & $0.10 \pm 0.02$ & $0.07 \pm 0.01$ & $0.13 \pm 0.05$ & $0.02 \pm 0.01$    & $0.41 \pm 0.06$ & $0.19 \pm 0.03$ & $0.10 \pm 0.02$ & $0.17 \pm 0.05$  & $0.02 \pm 0.01$  \\
Background rate [Mton $\cdot$ year] & $1.9 \pm 0.4$ & $2.0 \pm 0.4$ & $2.2 \pm 0.3$ & $0.6 \pm 0.2$ & $0.7 \pm 0.3$  & $4.5 \pm 0.6$ & $3.9 \pm 0.6$ & $3.4 \pm 0.6$ & $0.9 \pm 0.3$ & $0.7 \pm 0.4$ \\
Background syst. uncertainty [\%] & $40.2 \pm 8.7$  & $46.6 \pm 9.9$ & $33.4 \pm 6.8$ & $45.7 \pm 17.2$ & $53.1 \pm 21.0$  & $22.7 \pm 3.0$ & $29.1 \pm 4.1$ & $26.3 \pm 4.1$ & $30.2 \pm 9.1$ & $28.3 \pm 8.9$   \\
Data candidates & 0 & 1 & 0 & 0 & 0 &  0 & 0 & 0 & 1 & 0 \\
\midrule
Lifetime limit & \multicolumn{5}{c||}{$7.2 \times 10^{33}$ years (90\% C.L.)} & \multicolumn{5}{c}{$4.5 \times 10^{33}$ years (90\% C.L.)} \\
\bottomrule
\end{tabular}
}
\end{table*}

  Figure~\ref{fig:fig_totM} shows the reconstructed invariant mass distributions for both decay modes, after applying all event selection criteria, except for the reconstructed invariant mass cut itself, for signal MC, atm.-$\nu$ background MC, and data. 
  Decays of free protons are well reconstructed around the proton mass (938 MeV/$c^2$) in the MC simulation, while those of bound protons have a tail  toward lower invariant masses due to energy loss from neutral pion interactions within the oxygen nucleus. 
  Figure~\ref{fig:fig_totP} shows the total momentum distributions for both decay modes after applying all event selection criteria, except for the total momentum cut itself, for signal MC, atm.-$\nu$ background MC, and data. 
  The signal MC simulation for free protons from the $p\rightarrow e^+\pi^0\pi^0$ decay showed that the fraction of the events above the cut value of 200 MeV/$c$ was negligibly small. 
 The bump above 200 MeV/$c$ in the reconstructed momentum distribution of free protons from the $p\rightarrow \mu^+\pi^0\pi^0$ decay mode is caused by events where the muon is not reconstructed and one of the four gammas from the two $\pi^0$ decays is misidentified as a muon. 
 This occurs when the true muon momentum is above the Cherenkov threshold (118 MeV/$c$) but still relatively low, where reconstruction becomes difficult. 
 Figure~\ref{fig:fig_scatter} shows the correlation between the invariant mass and the total momentum for signal MC, atm.-$\nu$ background MC, and all data from SK-I to SK-V after applying all selection cuts except those on the plotted variables.   
 Figure~\ref{fig:fig_evtreduction} shows the signal detection efficiencies and expected number of atm.-$\nu$ background events along the event selections for the signal and background MCs, respectively. 
 The dominant reduction in background events comes from the reconstructed invariant mass and total momentum cuts (C6 and C7) while keeping the signal detection efficiencies relatively high. Table~\ref{tab:tab_sigeff} shows the signal detection efficiencies for each detector phase. 
 For both searches, the signal detection efficiencies in SK-II are lower than those in other detector phases because of the degradation of the particle identification performance during this period with only half the usual SK photocathode coverage. 
 For the $p\rightarrow \mu^+\pi^0\pi^0$ decay mode, the signal detection efficiencies in SK-IV and SK-V were higher than SK-I to SK-III due to the higher Michel electron tagging efficiency achieved with the QBEE and data acquisition systsem.
    
   To understand the background composition, the breakdown of the neutrino interaction mode-composition of atm.-$\nu$ background MC events that survived the event selection is summarized in Table~\ref{tab:tab_bkgint}. 
   In both proton decay searches, neutrino interactions of charged-current ($CC$) pion production interactions and neutral-current ($NC$) multi-pion productions were dominant. 
   The charged-current quasi-elastic ($CCQE$) events are not negligible, because the recoil nucleons from higher energy neutrinos can produce a $\pi^0$ in water. 
 For the $p\rightarrow \mu^+\pi^0\pi^0$ search, charged-current single $\pi^+$ ($CC1\pi^+$) production is one of the dominant modes due to $\pi^+$ to $\pi^0$ conversion via charge exchange in oxygen nuclei and to the $\pi^+$ being reconstructed as a non-shower ring, either directly or via $\pi^+$ decay. 
 All remaining atm.-$\nu$ background MC events were inspected with the SK event display. All events looked reasonable and no mis-reconstructed events were found.


\section{\label{sec:lab5}Search results}
 The distributions of the cut variables and the event reductions at each cut stage (C1-C7) agree well between data and the atm.-$\nu$ MC events.

 Table~\ref{tab:tab_summary} summarizes the exposure, the signal detection efficiency and its systematic uncertainty, the expected number of atm.-$\nu$ background events and its systematic uncertainty, and the number of proton decay candidates remaining in the SK data after all event selections. 
 One data candidate was observed for the $p\rightarrow e^+\pi^0\pi^0$ search in SK-II and one for the $p\rightarrow \mu^+\pi^0\pi^0$ search in SK-IV. 
 The Poisson probability of observing one or more events given the expected atm.-$\nu$ background is 9.5\% for $p \rightarrow e^+ \pi^0 \pi^0$ in SK-II and 15.6\% for $p \rightarrow \mu^+ \pi^0 \pi^0$ in SK-IV. 
 The expected number of atm.-$\nu$ background events for SK-I to SK-V exposures are 0.49 and 0.90, and the Poisson probabilities of observing one or more candidate events given the atm.-$\nu$ background expectations are 38.7\% and 59.4\%, respectively.  
 Therefore, we conclude that no significant excess of data events beyond the expected number of atm.-$\nu$ background events is observed.
 
  In the atm.-$\nu$ background events, the dependence of the selection efficiency on the detector phase (SK-I to SK-III) is expected to be comparable to that of the signal. 
  From SK-IV onward the new electronics allowed for improved Michel electron tagging and enabled neutron tagging, which reduced the BG by more than 50\%.
  Thus, the expected number of atm.-$\nu$ background events in SK-IV and SK-V are significantly lower than those in SK-I to SK-III.
  
  The reconstructed decay patterns overlayed over the two candidate SK events' hit patterns in SK's inner and outer detectors are shown in Figure~\ref{fig:fig_evtdisp}.
  For $p\rightarrow e^+\pi^0\pi^0$, the reconstructed vertex is located 6.5 m away from the nearest ID wall, the number of reconstructed rings is 4, the reconstructed invariant mass is 820 MeV/$c^{2}$ and the total momentum is 142 MeV/$c$. 
  And for $p\rightarrow \mu^+\pi^0\pi^0$, the reconstructed vertex is located 3.6 m away from the nearest ID wall, the number of reconstructed rings is 5, the reconstructed invariant mass is 946 MeV/$c^{2}$ and the total momentum is 177 MeV/$c$. 
  In each decay mode, the candidate event is located near the boundary of the signal region shown in Figures~\ref{fig:fig_totM} and \ref{fig:fig_totP}.
 
\begin{table}[hb!]
    \centering
    \renewcommand{\arraystretch}{1.2}
    \caption{\footnotesize The summary of systematic uncertainties of the signal detection efficiencies which are averaged over the total livetime from SK-I to SK-V for $p\rightarrow e^+\pi^0\pi^0$ and $p\rightarrow \mu^+\pi^0\pi^0$, respectively. The upper part lists physics model uncertainties, while the lower part lists detector-related uncertainties from event reconstruction. The systematic uncertainty of the number of tagged neutrons became accessible from SK-IV onward. The error shown for each contribution to these uncertainties is the statistical error of the MC statistics.}
    \label{tab:tab_systsig} 

    {\small
    \resizebox{0.8\columnwidth}{!}{%
    \begin{tabular}{lcc}
        \toprule
        & $p \to e^+ \pi^0 \pi^0 [\%] $ & $p \to \mu^+ \pi^0 \pi^0 [\%] $ \\
        \midrule
        \textbf{Physics model} \\
        \hspace{1.0em}Correlated decay & $2.1 \pm 0.4$ & $2.6 \pm 0.2$ \\
         \hspace{1.0em}Fermi momentum & $2.6 \pm 0.4$ & $3.0 \pm 0.2$ \\
         \hspace{1.0em}$\pi$-FSI & $8.4 \pm 0.2$ & $10.8 \pm 0.4$ \\
        \textbf{Total (Physics model)} & $\mathbf{9.0 \pm 0.6}$ & $\mathbf{11.5 \pm 0.4}$ \\
        \midrule
        \textbf{Event reconstruction} \\
         \hspace{1.0em}Fiducial volume & $1.8 \pm 0.1$ & $1.8 \pm 0.1$ \\
         \hspace{1.0em}Number of rings & $1.1 \pm 0.2$ & $0.1 \pm 0.1$ \\
         \hspace{1.0em}Particle identification & $4.0 \pm 0.5$ & $3.4 \pm 0.1$ \\
         \hspace{1.0em}Number of Michel electrons & $2.1 \pm 0.2$ & $2.0 \pm 0.2$ \\
         \hspace{1.0em}Energy Scale & & \\
        \hspace{2em}Absolute \& Time variation & $0.6 \pm 0.01$ & $1.1 \pm 0.01$ \\
        \hspace{2em}Non-uniformity & $0.6 \pm 0.01$ & $0.8 \pm 0.01$ \\
         \hspace{1.0em}Number of tagged neutrons & $0.2 \pm 0.01$ & $0.2 \pm 0.01$ \\
        \textbf{Total (Reconstruction)} & $\mathbf{5.1 \pm 0.6}$ & $\mathbf{4.6 \pm 0.2}$ \\
        \midrule
        \textbf{Total systematic uncertainty} & $\mathbf{10.3 \pm 0.8}$ & $\mathbf{12.4 \pm 0.4}$ \\
        \bottomrule
    \end{tabular}%
    }}
\end{table}

\begin{table}[hb!]
    \centering
    \renewcommand{\arraystretch}{1.2}
    \caption{\footnotesize The summary of systematic uncertainties of the expected number of atm.-$\nu$ background events, averaged over the total livetime from SK-I to SK-V, for $p\rightarrow e^+\pi^0\pi^0$ and $p\rightarrow \mu^+\pi^0\pi^0$, respectively. The upper part lists physics model uncertainties, while the lower part lists detector-related uncertainties from event reconstruction. The systematic uncertainty of the number of tagged neutrons became accessible from SK-IV onward. The error shown for each contribution to these uncertainties is the statistical error of the MC statistics.}
    \label{tab:tab_systbkg} 

    {\small
    \resizebox{0.8\columnwidth}{!}{%
    \begin{tabular}{lcc}
        \toprule
        & $p \to e^+ \pi^0 \pi^0 [\%] $ & $p \to \mu^+ \pi^0 \pi^0 [\%] $ \\
        \midrule
        \textbf{Physics Model} \\
         \hspace{1.0em}$\pi$-FSI \& SI & $23.3 \pm 7.5$ & $16.1 \pm 4.3$ \\
         \hspace{1.0em}Neutrino Flux & $6.8 \pm 2.1$ & $7.4 \pm 1.9$ \\
         \hspace{1.0em}Neutrino Interaction & $19.5 \pm 6.4$ & $17.1 \pm 4.6$ \\
        \textbf{Total (Physics Model)} & $\mathbf{31.2 \pm 10.1}$ & $\mathbf{25.0 \pm 6.7}$ \\
        \midrule
        \textbf{Event Reconstruction} \\
         \hspace{1.0em}Fiducial Volume & $1.8 \pm 0.1$ & $1.8 \pm 0.1$ \\
         \hspace{1.0em}Number of Rings & $11.3 \pm 3.9$ & $0.01 \pm 0.01$ \\
         \hspace{1.0em}Particle identification & $7.8 \pm 1.7$ & $6.5 \pm 1.3$ \\
         \hspace{1.0em}Number of Michel Electrons & $2.1 \pm 0.2$ & $2.0 \pm 0.1$ \\
         \hspace{1.0em}Energy Scale \\
        \quad Absolute \& Time Variation & $26.6 \pm 9.0$ & $7.2 \pm 1.5$ \\
        \quad Non-uniformity & $5.4 \pm 1.9$ & $6.2 \pm 1.5$ \\
         \hspace{1.0em}Number of Tagged Neutrons & $10.7 \pm 4.0$ & $6.5 \pm 2.0$ \\
        \textbf{Total (Reconstruction)} & $\mathbf{31.6 \pm 10.5}$ & $\mathbf{12.8 \pm 2.5}$ \\
        \midrule
        \textbf{Total systematic uncertainty} & $\mathbf{44.4 \pm 14.6}$ & $\mathbf{28.1 \pm 7.1}$ \\
        \bottomrule
    \end{tabular}%
    }}
\end{table}

 
 \section{\label{sec:lab6}Systematic uncertainty estimation}
 
 The total systematic uncertainties of signal detection efficiency and expected number of atm.-$\nu$ background events are summarized in Tables~\ref{tab:tab_systsig} and \ref{tab:tab_systbkg}, respectively. 
 These tables present livetime-weighted averages over SK-I to SK-V, while for the results presented in this paper, all systematic uncertainties are evaluated and applied separately for each SK phase in the lifetime limit calculation.
 The sources of systematic uncertainties and the systematic uncertainty estimation methods used in this analysis are the same as those used in SK's latest published $p\rightarrow \ell^+\pi^0$ analysis~\cite{Takenaka2020}.
 
 For both decay modes, the main systematic uncertainty for signal detection efficiency came from the $\pi$-FSI in oxygen nuclei. 
 This uncertainty is larger for the muonic decay channel as the muonic channel contains approximately 20\% more bound proton decays. 
 Fermi motion leads to a broader smearing of the pion kinematics, and the higher Cherenkov threshold for muons increases the impact of these effects on the signal efficiency.
  In the signal, the neutral pion promptly decays into two gammas before interacting with water, so the pion interactions in water (secondary interactions, SI) were not taken into account.
    Apart from the $\pi$-FSI uncertainties, no significant differences were observed in the systematic uncertainties of the signal detection efficiencies and the atm.-$\nu$ background event numbers of the $p\rightarrow e^+\pi^0\pi^0$ and $p\rightarrow \mu^+\pi^0\pi^0$ modes, and they are smaller than their statistical errors.
 The main uncertainties in the atm.-$\nu$ background event numbers come from $\pi$-FSI and SI, neutrino interactions, and energy scale.


\section{\label{sec:lab7}Lifetime limit calculation}
 The number of candidate events in each mode is consistent with the expected atm.-$\nu$ background as described in Section~\ref{sec:lab5}. 
 Since no significant excess of decay candidates was observed in either decay mode, a lower limit on the partial proton lifetime for each mode was calculated using the same Bayesian method as was used in Refs.~\cite{Amsler2008, Roe2000, Matsumoto2022}. 

  The signal detection efficiencies, the expected atm.-$\nu$ background events, and the systematic uncertainties for each detector phase are used as inputs for the calculation, and each detector phase is treated independently.
 A combined overall lifetime limit for each of the two decay modes is then determined by combining the independent measurements from the SK-I through SK-V phases, resulting in five signal boxes (C6–C7 region in each SK phase, defined by the single-box method introduced in Section~\ref{sec:lab4}). 
The posterior PDF for the proton decay rate ($P(\Gamma \mid n_i)$) is expressed as follows:
\begin{equation}
\begin{aligned}
P(\Gamma \mid n_i) &= \iiint \frac{e^{-(\Gamma\lambda_i\epsilon_i + b_i)}
(\Gamma\lambda_i\epsilon_i + b_i)^{n_i}}{n_i!} \\
&\quad \times P(\Gamma)\,P(\lambda_i)\,P(\epsilon_i)\,P(b_i)\,
d\lambda_i\,d\epsilon_i\,db_i,
\end{aligned}
\end{equation}
 where $n_{i}$ is the number of data candidate events, $i$ the index of each SK phase, $\lambda_{i}$ that phase's exposure, $\epsilon_{i}$ its signal detection efficiency and $b_{i}$ the number of expected atm.-$\nu$ background events. 
 The prior PDF for the proton decay rate $P(\Gamma)$ is assumed to be uniform, with $P(\Gamma)$ being 1 for $\Gamma \geq 0 $ and 0 otherwise. $P(\lambda_{i})$, $P(\epsilon_{i})$ and $P(b_{i})$ stand for the prior probability distributions for the detector exposure, the signal detection efficiency, and the expected number of atm.-$\nu$ backgrounds, respectively, which are assumed to be Gaussian distributions whose means correspond to the exposure, the signal efficiency, and the expected background, respectively, and whose widths correspond to the respective systematic uncertainties. 
 The systematic uncertainty of the detector exposure is estimated to be less than 1\% for all SK phases~\cite{Takenaka2020}.

The combined upper limit for the proton decay rate for the five detector phases at a given confidence level (C.L.) is:
\begin{equation}
\mathrm{C.L.} =
\int_{0}^{\Gamma_{\text{limit}}}
\prod_{i=1}^{5} P\!\left(\Gamma \mid n_i\right)\, d\Gamma.
\end{equation}
The lower limit on the partial lifetime for proton decay is given by:
 
\begin{equation}
\frac{\tau_{\text{limit}}}{B} = \frac{1}{\Gamma_{\text{limit}}}.
\end{equation}
where B represents the branching ratio of each proton decay mode.

The expected lifetime limits were $8.9 \times 10^{33}$ years and $5.7 \times 10^{33}$ years for $p \rightarrow e^+ \pi^0 \pi^0$ and $p \rightarrow \mu^+ \pi^0 \pi^0$, respectively, obtained from the mean of pseudo-experiments under the background-only hypothesis. 
The new lifetime limits derived from SK-I to SK-V data, in which one candidate event was observed in each decay mode and both events were consistent with the expected atm.-$\nu$ background, were found to be $\tau/B(p \rightarrow e^+ \pi^0 \pi^0) > 7.2 \times 10^{33}$ years and $\tau/B(p \rightarrow \mu^+ \pi^0 \pi^0) > 4.5 \times 10^{33}$ years at 90\% C.L.


\section{\label{sec:lab8}Conclusion}
 We searched for proton decay into a charged anti-lepton and two neutral pions using the data of all pure-water phases of Super-Kamiokande I-V. 
 No significant event excess above the expected atm.-$\nu$ background for either decay modes was found in 0.401 megaton-years exposure. 
 
 Lower limits on the partial lifetimes of $7.2\times10^{33}$ years for $p\rightarrow e^+\pi^0\pi^0$ and $4.5\times10^{33}$ years for $p\rightarrow \mu^+\pi^0\pi^0$ are obtained at the 90\% confidence level. 
 These limits are approximately 50 times higher for both decay modes compared to the results from the IMB-3 experiment and are the most stringent limits that exist in these modes.

\begin{acknowledgments}
\input{SK-paper-acknowledgements-20251121}

\appendix

\nocite{*}
\bibliographystyle{apsrev4-2}
\setcitestyle{numbers, compress, etalmode=1, maxnames=3}
\bibliography{PDK_lp2pi0}
\end{acknowledgments}
\end{document}

%% file: authors_20230602_20260407-orcid.tex
\newcommand{\AFFicrr}{\affiliation{Kamioka Observatory, Institute for Cosmic Ray Research, University of Tokyo, Kamioka, Gifu 506-1205, Japan}}
\newcommand{\AFFkashiwa}{\affiliation{Research Center for Cosmic Neutrinos, Institute for Cosmic Ray Research, University of Tokyo, Kashiwa, Chiba 277-8582, Japan}}
\newcommand{\AFFipmu}{\affiliation{Kavli Institute for the Physics and
Mathematics of the Universe (WPI), The University of Tokyo Institutes for Advanced Study,
University of Tokyo, Kashiwa, Chiba 277-8583, Japan }}
\newcommand{\AFFmad}{\affiliation{Department of Theoretical Physics, University Autonoma Madrid, 28049 Madrid, Spain}}
\newcommand{\AFFubc}{\affiliation{Department of Physics and Astronomy, University of British Columbia, Vancouver, BC, V6T1Z4, Canada}}
\newcommand{\AFFbu}{\affiliation{Department of Physics, Boston University, Boston, MA 02215, USA}}
\newcommand{\AFFuci}{\affiliation{Department of Physics and Astronomy, University of California, Irvine, Irvine, CA 92697-4575, USA }}
\newcommand{\AFFucd}{\affiliation{Crocker Nuclear Lab, University of California Davis, Davis, CA 95616, USA}}
\newcommand{\AFFcsu}{\affiliation{Department of Physics, California State University, Dominguez Hills, Carson, CA 90747, USA}}
\newcommand{\AFFcnm}{\affiliation{Institute for Universe and Elementary Particles, Chonnam National University, Gwangju 61186, Korea}}
\newcommand{\AFFduke}{\affiliation{Department of Physics, Duke University, Durham NC 27708, USA}}
\newcommand{\AFFgifu}{\affiliation{Department of Physics, Gifu University, Gifu, Gifu 501-1193, Japan}}
\newcommand{\AFFgist}{\affiliation{GIST College, Gwangju Institute of Science and Technology, Gwangju 500-712, Korea}}
\newcommand{\AFFuh}{\affiliation{Department of Physics and Astronomy, University of Hawaii, Honolulu, HI 96822, USA}}
\newcommand{\AFFicl}{\affiliation{Department of Physics, Imperial College London , London, SW7 2AZ, United Kingdom }}
\newcommand{\AFFkek}{\affiliation{High Energy Accelerator Research Organization (KEK), Tsukuba, Ibaraki 305-0801, Japan }}
\newcommand{\AFFkobe}{\affiliation{Department of Physics, Kobe University, Kobe, Hyogo 657-8501, Japan}}
\newcommand{\AFFkyoto}{\affiliation{Department of Physics, Kyoto University, Kyoto, Kyoto 606-8502, Japan}}
\newcommand{\AFFliv}{\affiliation{Department of Physics, University of Liverpool, Liverpool, L69 7ZE, United Kingdom}}
\newcommand{\AFFmiyagi}{\affiliation{Department of Physics, Miyagi University of Education, Sendai, Miyagi 980-0845, Japan}}
\newcommand{\AFFnagoya}{\affiliation{Institute for Space-Earth Environmental Research, Nagoya University, Nagoya, Aichi 464-8602, Japan}}
\newcommand{\AFFkmi}{\affiliation{Kobayashi-Maskawa Institute for the Origin of Particles and the Universe, Nagoya University, Nagoya, Aichi 464-8602, Japan}}
\newcommand{\AFFpol}{\affiliation{National Centre For Nuclear Research, 02-093 Warsaw, Poland}}
\newcommand{\AFFsuny}{\affiliation{Department of Physics and Astronomy, State University of New York at Stony Brook, NY 11794-3800, USA}}
\newcommand{\AFFokayama}{\affiliation{Department of Physics, Okayama University, Okayama, Okayama 700-8530, Japan }}
\newcommand{\AFFosaka}{\affiliation{Department of Physics, Osaka University, Toyonaka, Osaka 560-0043, Japan}}
\newcommand{\AFFox}{\affiliation{Department of Physics, Oxford University, Oxford, OX1 3PU, United Kingdom}}
\newcommand{\AFFqmul}{\affiliation{School of Physics and Astronomy, Queen Mary University of London, London, E1 4NS, United Kingdom}}
\newcommand{\AFFregina}{\affiliation{Department of Physics, University of Regina, 3737 Wascana Parkway, Regina, SK, S4SOA2, Canada}}
\newcommand{\AFFseoul}{\affiliation{Department of Physics and Astronomy, Seoul National University, Seoul 151-742, Korea}}
\newcommand{\AFFsheff}{\affiliation{School of Mathematical and Physical Science, University of Sheffield, S3 7RH, Sheffield, United Kingdom}}
\newcommand{\AFFshizuokasc}{\affiliation{Department of Informatics in
Social Welfare, Shizuoka University of Welfare, Yaizu, Shizuoka, 425-8611, Japan}}
\newcommand{\AFFstfc}{\affiliation{STFC, Rutherford Appleton Laboratory, Harwell Oxford, and Daresbury Laboratory, Warrington, OX11 0QX, United Kingdom}}
\newcommand{\AFFskk}{\affiliation{Department of Physics, Sungkyunkwan University, Suwon 440-746, Korea}}
\newcommand{\AFFtodai}{\affiliation{Department of Physics, University of Tokyo, Bunkyo, Tokyo 113-0033, Japan }}
\newcommand{\AFFtit}{\affiliation{Department of Physics, Institute of Science Tokyo, Meguro, Tokyo 152-8551, Japan }}
\newcommand{\AFFtus}{\affiliation{Department of Physics and Astronomy, Faculty of Science and Technology, Tokyo University of Science, Noda, Chiba 278-8510, Japan }}
\newcommand{\AFFtriumf}{\affiliation{TRIUMF, 4004 Wesbrook Mall, Vancouver, BC, V6T2A3, Canada }}
\newcommand{\AFFtokai}{\affiliation{Department of Physics, Tokai University, Hiratsuka, Kanagawa 259-1292, Japan}}
\newcommand{\AFFtsinghua}{\affiliation{Department of Engineering Physics, Tsinghua University, Beijing, 100084, China}}
\newcommand{\AFFynu}{\affiliation{Department of Physics, Yokohama National University, Yokohama, Kanagawa, 240-8501, Japan}}
\newcommand{\AFFllr}{\affiliation{Ecole Polytechnique, IN2P3-CNRS, Laboratoire Leprince-Ringuet, F-91120 Palaiseau, France }}
\newcommand{\AFFbari}{\affiliation{ Dipartimento Interuniversitario di Fisica, INFN Sezione di Bari and Universit\`a e Politecnico di Bari, I-70125, Bari, Italy}}
\newcommand{\AFFnapoli}{\affiliation{Dipartimento di Fisica, INFN Sezione di Napoli and Universit\`a di Napoli, I-80126, Napoli, Italy}}
\newcommand{\AFFroma}{\affiliation{INFN Sezione di Roma and Universit\`a di Roma ``La Sapienza'', I-00185, Roma, Italy}}
\newcommand{\AFFpadova}{\affiliation{Dipartimento di Fisica, INFN Sezione di Padova and Universit\`a di Padova, I-35131, Padova, Italy}}
\newcommand{\AFFkeio}{\affiliation{Department of Physics, Keio University, Yokohama, Kanagawa, 223-8522, Japan}}
\newcommand{\AFFwinnipeg}{\affiliation{Department of Physics, University of Winnipeg, MB R3J 3L8, Canada }}
\newcommand{\AFFkcl}{\affiliation{Department of Physics, King's College London, London, WC2R 2LS, UK }}
\newcommand{\AFFwarwick}{\affiliation{Department of Physics, University of Warwick, Coventry, CV4 7AL, UK }}
\newcommand{\AFFral}{\affiliation{Rutherford Appleton Laboratory, Harwell, Oxford, OX11 0QX, UK }}
\newcommand{\AFFwu}{\affiliation{Faculty of Physics, University of Warsaw, Warsaw, 02-093, Poland }}
\newcommand{\AFFbcit}{\affiliation{Department of Physics, British Columbia Institute of Technology, Burnaby, BC, V5G 3H2, Canada }}
\newcommand{\AFFtohoku}{\affiliation{Department of Physics, Faculty of Science, Tohoku University, Sendai, Miyagi, 980-8578, Japan }}
\newcommand{\AFFicise}{\affiliation{Institute For Interdisciplinary Research in Science and Education, ICISE, Quy Nhon, 55121, Vietnam }}
\newcommand{\AFFilance}{\affiliation{ILANCE, CNRS - University of Tokyo International Research Laboratory, Kashiwa, Chiba 277-8582, Japan}}
\newcommand{\AFFibs}{\affiliation{Center for Underground Physics, Institute for Basic Science (IBS), Daejeon, 34126, Korea}}
\newcommand{\AFFglasgow}{\affiliation{School of Physics and Astronomy, University of Glasgow, Glasgow, Scotland, G12 8QQ, United Kingdom}}
\newcommand{\AFFoecu}{\affiliation{Media Communication Center, Osaka Electro-Communication University, Neyagawa, Osaka, 572-8530, Japan}}
\newcommand{\AFFminn}{\affiliation{School of Physics and Astronomy, University of Minnesota, Minneapolis, MN  55455, USA}}
\newcommand{\AFFsilesia}{\affiliation{August Che\l{}kowski Institute of Physics, University of Silesia in Katowice, 75 Pu\l{}ku Piechoty 1, 41-500 Chorz\'{o}w, Poland}}
\newcommand{\AFFtoyama}{\affiliation{Faculty of Science, University of Toyama, Toyama City, Toyama 930-8555, Japan}}
\newcommand{\AFFbmcc}{\affiliation{Science Department, Borough of Manhattan Community College / City University of New York, New York, New York, 1007, USA.}}
\newcommand{\AFFnumazu}{\affiliation{National Institute of Technology, Numazu College, Numazu, Shizuoka 410-8501, Japan}}

\AFFicrr
\AFFkashiwa
\AFFmad
\AFFbmcc
\AFFbu
\AFFbcit
\AFFuci
\AFFucd
\AFFcsu
\AFFcnm
\AFFduke
\AFFllr
\AFFgifu
\AFFgist
\AFFglasgow
\AFFuh
\AFFibs
\AFFicise
\AFFicl
\AFFbari
\AFFnapoli
\AFFpadova
\AFFroma
\AFFilance
\AFFkeio
\AFFkek
\AFFkcl
\AFFkobe
\AFFkyoto
\AFFliv
\AFFminn
\AFFmiyagi
\AFFnagoya
\AFFkmi
\AFFpol
\AFFnumazu
\AFFsuny
\AFFokayama
\AFFoecu
\AFFox
\AFFral
\AFFseoul
\AFFsheff
\AFFshizuokasc
\AFFsilesia
\AFFstfc
\AFFskk
\AFFtohoku
\AFFtokai
\AFFtodai
\AFFipmu
\AFFtit
\AFFtus
\AFFtoyama
\AFFtriumf
\AFFtsinghua
\AFFwu
\AFFwarwick
\AFFwinnipeg
\AFFynu

\author[0009-0000-9620-788X]{K.~Abe}
\AFFicrr
\AFFipmu
\author[0000-0002-2110-5130]{S.~Abe}
\AFFicrr
\AFFtodai
\author[0000-0001-6440-933X]{Y.~Asaoka}
\AFFicrr
\AFFipmu
\author[0000-0003-3273-946X]{M.~Harada}
\AFFicrr
\author[0000-0002-8683-5038]{Y.~Hayato}
\AFFicrr
\AFFipmu
\author[0000-0003-1229-9452]{K.~Hiraide}
\AFFicrr
\AFFipmu
\author{T.~H.~Hung}
\AFFicrr
\author[0000-0002-8766-3629]{K.~Hosokawa}
\AFFicrr
\author[0000-0002-7791-5044]{K.~Ieki}
\author[0000-0002-4177-5828]{M.~Ikeda}
\AFFicrr
\AFFipmu
% ...


\collaboration{The Super-Kamiokande Collaboration}
\noaffiliation

%% file: SK-paper-acknowledgements-20251121.tex




We gratefully acknowledge the cooperation of the Kamioka Mining and Smelting Company. The Super-Kamiokande experiment has been built and operated from funding by the Japanese Ministry of Education, Culture, Sports, Science and Technology; the U.S. Department of Energy; and the U.S. National Science Foundation. Some of us have been supported by funds from the National Research Foundation of Korea (NRF-2009-0083526, NRF-2022R1A5A1030700, NRF-2022R1A3B1078756, RS-2025-00514948) funded by the Ministry of Science, Information and Communication Technology (ICT); the Institute for Basic Science (IBS-R016-Y2); and the Ministry of Education (2018R1D1A1B07049158, 2021R1I1A1A01042256, RS-2024-00442775); the Japan Society for the Promotion of Science; the National Natural Science Foundation of China (Grants No. 12375100 and 12521007); the Spanish Ministry of Science, Universities and Innovation (grant PID2021-124050NB-C31); the Natural Sciences and Engineering Research Council (NSERC) of Canada; the Scinet and Digital Research of Alliance Canada; the National Science Centre (UMO-2018/30/E/ST2/00441 and UMO-2022/46/E/ST2/00336) and the Ministry of  Science and Higher Education (2023/WK/04), Poland; the Science and Technology Facilities Council (STFC) and Grid for Particle Physics (GridPP), UK; the European Union’s Horizon 2020 Research and Innovation Programme H2020-MSCA-RISE-2018 JENNIFER2 grant agreement no.822070, H2020-MSCA-RISE-2019 SK2HK grant agreement no. 872549; and European Union's Next Generation EU/PRTR  grant CA3/RSUE2021-00559; the National Institute for Nuclear Physics (INFN), Italy.

%% file: PDK_lp2pi0.bbl
\begin{thebibliography}{29}%
\makeatletter
\providecommand \@ifxundefined [1]{%
 \@ifx{#1\undefined}
}%
\providecommand \@ifnum [1]{%
 \ifnum #1\expandafter \@firstoftwo
 \else \expandafter \@secondoftwo
 \fi
}%
\providecommand \@ifx [1]{%
 \ifx #1\expandafter \@firstoftwo
 \else \expandafter \@secondoftwo
 \fi
}%
\providecommand \natexlab [1]{#1}%
\providecommand \enquote  [1]{``#1''}%
\providecommand \bibnamefont  [1]{#1}%
\providecommand \bibfnamefont [1]{#1}%
\providecommand \citenamefont [1]{#1}%
\providecommand \href@noop [0]{\@secondoftwo}%
\providecommand \href [0]{\begingroup \@sanitize@url \@href}%
\providecommand \@href[1]{\@@startlink{#1}\@@href}%
\providecommand \@@href[1]{\endgroup#1\@@endlink}%
\providecommand \@sanitize@url [0]{\catcode `\\12\catcode `\$12\catcode
  `\&12\catcode `\#12\catcode `\^12\catcode `\_12\catcode `\%12\relax}%
\providecommand \@@startlink[1]{}%
\providecommand \@@endlink[0]{}%
\providecommand \url  [0]{\begingroup\@sanitize@url \@url }%
\providecommand \@url [1]{\endgroup\@href {#1}{\urlprefix }}%
\providecommand \urlprefix  [0]{URL }%
\providecommand \Eprint [0]{\href }%
\providecommand \doibase [0]{https://doi.org/}%
\providecommand \selectlanguage [0]{\@gobble}%
\providecommand \bibinfo  [0]{\@secondoftwo}%
\providecommand \bibfield  [0]{\@secondoftwo}%
\providecommand \translation [1]{[#1]}%
\providecommand \BibitemOpen [0]{}%
\providecommand \bibitemStop [0]{}%
\providecommand \bibitemNoStop [0]{.\EOS\space}%
\providecommand \EOS [0]{\spacefactor3000\relax}%
\providecommand \BibitemShut  [1]{\csname bibitem#1\endcsname}%
\let\auto@bib@innerbib\@empty
\bibitem [{\citenamefont {Workman}\ \emph {et~al.}(2022)\citenamefont {Workman}
  \emph {et~al.}}]{pdgGUT2023}%
  \BibitemOpen
  \bibfield  {author} {\bibinfo {author} {\bibfnamefont {R.~L.}\ \bibnamefont
  {Workman}} \emph {et~al.} (\bibinfo {collaboration} {Particle Data Group}),\
  }\bibfield  {title} {\bibinfo {title} {{Review of Particle Physics}},\ }\href
  {https://doi.org/10.1093/ptep/ptac097} {\bibfield  {journal} {\bibinfo
  {journal} {PTEP}\ }\textbf {\bibinfo {volume} {2022}},\ \bibinfo {pages}
  {083C01} (\bibinfo {year} {2022})}\BibitemShut {NoStop}%
\bibitem [{\citenamefont {Georgi}\ and\ \citenamefont
  {Glashow}(1974)}]{GeorgiSU(5)}%
  \BibitemOpen
  \bibfield  {author} {\bibinfo {author} {\bibfnamefont {H.}~\bibnamefont
  {Georgi}}\ and\ \bibinfo {author} {\bibfnamefont {S.~L.}\ \bibnamefont
  {Glashow}},\ }\bibfield  {title} {\bibinfo {title} {Unity of all
  elementary-particle forces},\ }\href
  {https://doi.org/10.1103/PhysRevLett.32.438} {\bibfield  {journal} {\bibinfo
  {journal} {Phys. Rev. Lett.}\ }\textbf {\bibinfo {volume} {32}},\ \bibinfo
  {pages} {438} (\bibinfo {year} {1974})}\BibitemShut {NoStop}%
\bibitem [{\citenamefont {Langacker}(1981)}]{LangackerGUT}%
  \BibitemOpen
  \bibfield  {author} {\bibinfo {author} {\bibfnamefont {P.}~\bibnamefont
  {Langacker}},\ }\bibfield  {title} {\bibinfo {title} {Grand unified theories
  and proton decay},\ }\href {https://doi.org/10.1016/0370-1573(81)90059-4}
  {\bibfield  {journal} {\bibinfo  {journal} {Phys. Rep.}\ }\textbf {\bibinfo
  {volume} {72}},\ \bibinfo {pages} {185} (\bibinfo {year} {1981})}\BibitemShut
  {NoStop}%
\bibitem [{\citenamefont {Ellis}\ \emph {et~al.}(2020)\citenamefont {Ellis},
  \citenamefont {Garcia}, \citenamefont {Nagata}, \citenamefont {Nanopoulos},\
  and\ \citenamefont {Olive}}]{EllisSU(5)}%
  \BibitemOpen
  \bibfield  {author} {\bibinfo {author} {\bibfnamefont {J.}~\bibnamefont
  {Ellis}}, \bibinfo {author} {\bibfnamefont {M.~A.~G.}\ \bibnamefont
  {Garcia}}, \bibinfo {author} {\bibfnamefont {N.}~\bibnamefont {Nagata}},
  \bibinfo {author} {\bibfnamefont {D.~V.}\ \bibnamefont {Nanopoulos}},\ and\
  \bibinfo {author} {\bibfnamefont {K.~A.}\ \bibnamefont {Olive}},\ }\bibfield
  {title} {\bibinfo {title} {Proton decay: Flipped vs. unflipped su(5)},\
  }\href {https://doi.org/10.1007/JHEP05(2020)021} {\bibfield  {journal}
  {\bibinfo  {journal} {J. High Energy Phys.}\ }\textbf {\bibinfo {volume}
  {05}},\ \bibinfo {pages} {021}}\BibitemShut {NoStop}%
\bibitem [{\citenamefont {Fritzsch}\ and\ \citenamefont
  {Minkowski}(1975)}]{FritzschSO(10)}%
  \BibitemOpen
  \bibfield  {author} {\bibinfo {author} {\bibfnamefont {H.}~\bibnamefont
  {Fritzsch}}\ and\ \bibinfo {author} {\bibfnamefont {P.}~\bibnamefont
  {Minkowski}},\ }\bibfield  {title} {\bibinfo {title} {Unified interactions of
  leptons and hadrons},\ }\href {https://doi.org/10.1016/0003-4916(75)90211-0}
  {\bibfield  {journal} {\bibinfo  {journal} {Ann. Phys. (N.Y.)}\ }\textbf
  {\bibinfo {volume} {93}},\ \bibinfo {pages} {193} (\bibinfo {year}
  {1975})}\BibitemShut {NoStop}%
\bibitem [{\citenamefont {Babu}\ and\ \citenamefont {Khan}(2015)}]{BabuSO(10)}%
  \BibitemOpen
  \bibfield  {author} {\bibinfo {author} {\bibfnamefont {K.~S.}\ \bibnamefont
  {Babu}}\ and\ \bibinfo {author} {\bibfnamefont {S.}~\bibnamefont {Khan}},\
  }\bibfield  {title} {\bibinfo {title} {Minimal nonsupersymmetric so(10)
  model: Gauge coupling unification, proton decay, and fermion masses},\ }\href
  {https://doi.org/10.1103/PhysRevD.92.075018} {\bibfield  {journal} {\bibinfo
  {journal} {Phys. Rev. D}\ }\textbf {\bibinfo {volume} {92}},\ \bibinfo
  {pages} {075018} (\bibinfo {year} {2015})}\BibitemShut {NoStop}%
\bibitem [{\citenamefont {Wise}\ \emph {et~al.}(1981)\citenamefont {Wise},
  \citenamefont {Blankenbecler},\ and\ \citenamefont {Abbott}}]{Wise3body}%
  \BibitemOpen
  \bibfield  {author} {\bibinfo {author} {\bibfnamefont {M.~B.}\ \bibnamefont
  {Wise}}, \bibinfo {author} {\bibfnamefont {R.}~\bibnamefont
  {Blankenbecler}},\ and\ \bibinfo {author} {\bibfnamefont {L.~F.}\
  \bibnamefont {Abbott}},\ }\bibfield  {title} {\bibinfo {title} {Three-body
  decays of the proton},\ }\href {https://doi.org/10.1103/PhysRevD.23.1591}
  {\bibfield  {journal} {\bibinfo  {journal} {Phys. Rev. D}\ }\textbf {\bibinfo
  {volume} {23}},\ \bibinfo {pages} {1591} (\bibinfo {year}
  {1981})}\BibitemShut {NoStop}%
\bibitem [{\citenamefont {Oset}(1988)}]{OsetMesonEx}%
  \BibitemOpen
  \bibfield  {author} {\bibinfo {author} {\bibfnamefont {E.}~\bibnamefont
  {Oset}},\ }\bibfield  {title} {\bibinfo {title} {Meson exchange currents in
  the p decay in nuclei},\ }\bibfield  {journal} {\bibinfo  {journal} {Nucl.
  Phys. B}\ }\textbf {\bibinfo {volume} {304}},\ \href
  {https://doi.org/10.1016/0550-3213(88)90656-6} {10.1016/0550-3213(88)90656-6}
  (\bibinfo {year} {1988})\BibitemShut {NoStop}%
\bibitem [{\citenamefont {McGrew}\ \emph {et~al.}(1999)\citenamefont {McGrew}
  \emph {et~al.}}]{IMB3}%
  \BibitemOpen
  \bibfield  {author} {\bibinfo {author} {\bibfnamefont {C.}~\bibnamefont
  {McGrew}} \emph {et~al.},\ }\bibfield  {title} {\bibinfo {title} {Search for
  nucleon decay using the imb-3 detector},\ }\href
  {https://doi.org/10.1103/PhysRevD.59.052004} {\bibfield  {journal} {\bibinfo
  {journal} {Phys. Rev. D}\ }\textbf {\bibinfo {volume} {59}},\ \bibinfo
  {pages} {052004} (\bibinfo {year} {1999})}\BibitemShut {NoStop}%
\bibitem [{\citenamefont {Takenaka}\ \emph {et~al.}(2020)\citenamefont
  {Takenaka} \emph {et~al.}}]{Takenaka2020}%
  \BibitemOpen
  \bibfield  {author} {\bibinfo {author} {\bibfnamefont {A.}~\bibnamefont
  {Takenaka}} \emph {et~al.} (\bibinfo {collaboration} {Super-Kamiokande
  Collaboration}),\ }\bibfield  {title} {\bibinfo {title} {Search for proton
  decay via $p \rightarrow e^{+} \pi^{0}$ and $p \rightarrow \mu^{+} \pi^{0}$
  with an enlarged fiducial volume in super-kamiokande i-iv},\ }\href
  {https://doi.org/10.1103/PhysRevD.102.112011} {\bibfield  {journal} {\bibinfo
   {journal} {Phys. Rev. D}\ }\textbf {\bibinfo {volume} {102}},\ \bibinfo
  {pages} {112011} (\bibinfo {year} {2020})}\BibitemShut {NoStop}%
\bibitem [{\citenamefont {Mine}(2025)}]{MineTextbook}%
  \BibitemOpen
  \bibfield  {author} {\bibinfo {author} {\bibfnamefont {S.}~\bibnamefont
  {Mine}},\ }\href {https://doi.org/10.1142/9789819801107_0008} {\bibinfo
  {title} {The super-kamiokande and other detectors: A case study of large
  volume cherenkov neutrino detectors}} (\bibinfo {year} {2025})\BibitemShut
  {NoStop}%
\bibitem [{\citenamefont {Fukuda}\ \emph {et~al.}(2003)\citenamefont {Fukuda}
  \emph {et~al.}}]{SK1NIM}%
  \BibitemOpen
  \bibfield  {author} {\bibinfo {author} {\bibfnamefont {S.}~\bibnamefont
  {Fukuda}} \emph {et~al.},\ }\bibfield  {title} {\bibinfo {title} {The
  super-kamiokande detector},\ }\href
  {https://doi.org/10.1016/S0168-9002(03)00425-X} {\bibfield  {journal}
  {\bibinfo  {journal} {Nucl. Instrum. Methods Phys. Res., Sect. A}\ }\textbf
  {\bibinfo {volume} {501}},\ \bibinfo {pages} {418} (\bibinfo {year}
  {2003})}\BibitemShut {NoStop}%
\bibitem [{\citenamefont {Abe}\ \emph {et~al.}(2014)\citenamefont {Abe} \emph
  {et~al.}}]{SK4NIM}%
  \BibitemOpen
  \bibfield  {author} {\bibinfo {author} {\bibfnamefont {K.}~\bibnamefont
  {Abe}} \emph {et~al.},\ }\bibfield  {title} {\bibinfo {title} {Calibration of
  the super-kamiokande detector},\ }\href
  {https://doi.org/10.1016/j.nima.2013.11.081} {\bibfield  {journal} {\bibinfo
  {journal} {Nucl. Instrum. Methods Phys. Res., Sect. A}\ }\textbf {\bibinfo
  {volume} {737}},\ \bibinfo {pages} {253} (\bibinfo {year}
  {2014})}\BibitemShut {NoStop}%
\bibitem [{\citenamefont {Abe}\ \emph {et~al.}(2022)\citenamefont {Abe} \emph
  {et~al.}}]{FirstSKGd}%
  \BibitemOpen
  \bibfield  {author} {\bibinfo {author} {\bibfnamefont {K.}~\bibnamefont
  {Abe}} \emph {et~al.} (\bibinfo {collaboration} {Super-Kamiokande
  Collaboration}),\ }\bibfield  {title} {\bibinfo {title} {First gadolinium
  loading to super-kamiokande},\ }\href
  {https://doi.org/10.1016/j.nima.2021.166248} {\bibfield  {journal} {\bibinfo
  {journal} {Nucl. Instrum. Methods Phys. Res., Sect. A}\ }\textbf {\bibinfo
  {volume} {1027}},\ \bibinfo {pages} {166248} (\bibinfo {year}
  {2022})}\BibitemShut {NoStop}%
\bibitem [{\citenamefont {Abe}\ \emph {et~al.}(2024)\citenamefont {Abe} \emph
  {et~al.}}]{SecondSKGd}%
  \BibitemOpen
  \bibfield  {author} {\bibinfo {author} {\bibfnamefont {K.}~\bibnamefont
  {Abe}} \emph {et~al.} (\bibinfo {collaboration} {Super-Kamiokande
  Collaboration}),\ }\bibfield  {title} {\bibinfo {title} {Second gadolinium
  loading to super-kamiokande},\ }\href
  {https://doi.org/10.1016/j.nima.2024.169480} {\bibfield  {journal} {\bibinfo
  {journal} {Nucl. Instrum. Methods Phys. Res., Sect. A}\ }\textbf {\bibinfo
  {volume} {1065}},\ \bibinfo {pages} {169480} (\bibinfo {year}
  {2024})}\BibitemShut {NoStop}%
\bibitem [{\citenamefont {Takenaka}(2020)}]{TakeThesis}%
  \BibitemOpen
  \bibfield  {author} {\bibinfo {author} {\bibfnamefont {A.}~\bibnamefont
  {Takenaka}},\ }\emph {\bibinfo {title} {Search for Proton Decay via $p
  \rightarrow e^{+} \pi^{0}$ and $p \rightarrow \mu^{+} \pi^{0}$ with an
  Enlarged Fiducial Mass of the Super-Kamiokande Detector}},\ \href
  {https://www-sk.icrr.u-tokyo.ac.jp/sk/_pdf/articles/thesis.Akira.Takenaka.20210118.submit.pdf}
  {Ph.D. thesis},\ \bibinfo  {school} {University of Tokyo} (\bibinfo {year}
  {2020}),\ \bibinfo {note} {phD Thesis}\BibitemShut {NoStop}%
\bibitem [{\citenamefont {Wester}(2023)}]{ThomasThesis}%
  \BibitemOpen
  \bibfield  {author} {\bibinfo {author} {\bibfnamefont {T.}~\bibnamefont
  {Wester}},\ }\emph {\bibinfo {title} {Discerning the neutrino mass ordering
  using atmospheric neutrinos in super-kamiokande I-V}},\ \href
  {https://www-sk.icrr.u-tokyo.ac.jp/sk/_pdf/articles/2023/twester_thesis_2023_embedded.pdf}
  {Ph.D. thesis},\ \bibinfo  {school} {Boston university} (\bibinfo {year}
  {2023}),\ \bibinfo {note} {phD Thesis}\BibitemShut {NoStop}%
\bibitem [{\citenamefont {Nakamura}\ \emph {et~al.}(1976)\citenamefont
  {Nakamura}, \citenamefont {Hiramatsu}, \citenamefont {Kamae}, \citenamefont
  {Muramatsu}, \citenamefont {Izutsu},\ and\ \citenamefont
  {Watase}}]{Fermimotion}%
  \BibitemOpen
  \bibfield  {author} {\bibinfo {author} {\bibfnamefont {K.}~\bibnamefont
  {Nakamura}}, \bibinfo {author} {\bibfnamefont {S.}~\bibnamefont {Hiramatsu}},
  \bibinfo {author} {\bibfnamefont {T.}~\bibnamefont {Kamae}}, \bibinfo
  {author} {\bibfnamefont {H.}~\bibnamefont {Muramatsu}}, \bibinfo {author}
  {\bibfnamefont {N.}~\bibnamefont {Izutsu}},\ and\ \bibinfo {author}
  {\bibfnamefont {Y.}~\bibnamefont {Watase}},\ }\bibfield  {title} {\bibinfo
  {title} {The reaction 12c(e, e$'$p) at 700 mev and dwia analysis},\ }\href
  {https://doi.org/10.1016/0375-9474(76)90539-X} {\bibfield  {journal}
  {\bibinfo  {journal} {Nuclear Physics A}\ }\textbf {\bibinfo {volume}
  {268}},\ \bibinfo {pages} {381} (\bibinfo {year} {1976})}\BibitemShut
  {NoStop}%
\bibitem [{\citenamefont {Mayer}\ and\ \citenamefont
  {Jensen}(1955)}]{BindingE}%
  \BibitemOpen
  \bibfield  {author} {\bibinfo {author} {\bibfnamefont {M.~G.}\ \bibnamefont
  {Mayer}}\ and\ \bibinfo {author} {\bibfnamefont {J.~H.~D.}\ \bibnamefont
  {Jensen}},\ }\href@noop {} {\emph {\bibinfo {title} {Elementary Theory of
  Nuclear Shell Structure}}}\ (\bibinfo  {publisher} {Wiley},\ \bibinfo
  {address} {New York},\ \bibinfo {year} {1955})\BibitemShut {NoStop}%
\bibitem [{\citenamefont {Yamazaki}\ and\ \citenamefont
  {Akaishi}(1999)}]{Corrdecay}%
  \BibitemOpen
  \bibfield  {author} {\bibinfo {author} {\bibfnamefont {T.}~\bibnamefont
  {Yamazaki}}\ and\ \bibinfo {author} {\bibfnamefont {Y.}~\bibnamefont
  {Akaishi}},\ }\bibfield  {title} {\bibinfo {title} {Nuclear medium effects on
  invariant mass spectra of hadrons decaying in nuclei},\ }\href
  {https://doi.org/10.1016/S0370-2693(99)00163-X} {\bibfield  {journal}
  {\bibinfo  {journal} {Phys. Lett. B}\ }\textbf {\bibinfo {volume} {453}},\
  \bibinfo {pages} {1} (\bibinfo {year} {1999})}\BibitemShut {NoStop}%
\bibitem [{\citenamefont {Hayato}(2002)}]{NEUTHayato}%
  \BibitemOpen
  \bibfield  {author} {\bibinfo {author} {\bibfnamefont {Y.}~\bibnamefont
  {Hayato}},\ }\bibfield  {title} {\bibinfo {title} {Neut},\ }\href
  {https://doi.org/10.1016/S0920-5632(02)01759-0} {\bibfield  {journal}
  {\bibinfo  {journal} {Nucl. Phys. B, Proc. Suppl.}\ }\textbf {\bibinfo
  {volume} {112}},\ \bibinfo {pages} {171} (\bibinfo {year}
  {2002})}\BibitemShut {NoStop}%
\bibitem [{\citenamefont {Mitsuka}(2007)}]{NEUTMitsuka}%
  \BibitemOpen
  \bibfield  {author} {\bibinfo {author} {\bibfnamefont {G.}~\bibnamefont
  {Mitsuka}},\ }\bibfield  {title} {\bibinfo {title} {Neut},\ }\href
  {https://doi.org/10.1063/1.2834480} {\bibfield  {journal} {\bibinfo
  {journal} {AIP Conf. Proc.}\ }\textbf {\bibinfo {volume} {967}},\ \bibinfo
  {pages} {208} (\bibinfo {year} {2007})}\BibitemShut {NoStop}%
\bibitem [{\citenamefont {Mitsuka}(2008)}]{AIP2008}%
  \BibitemOpen
  \bibfield  {author} {\bibinfo {author} {\bibfnamefont {G.}~\bibnamefont
  {Mitsuka}},\ }\bibfield  {title} {\bibinfo {title} {Aip conference
  proceedings},\ }\href {https://doi.org/10.1063/1.2898954} {\bibfield
  {journal} {\bibinfo  {journal} {AIP Conf. Proc.}\ }\textbf {\bibinfo {volume}
  {981}},\ \bibinfo {pages} {262} (\bibinfo {year} {2008})}\BibitemShut
  {NoStop}%
\bibitem [{\citenamefont {Honda}\ \emph {et~al.}(2011)\citenamefont {Honda},
  \citenamefont {Kajita}, \citenamefont {Kasahara},\ and\ \citenamefont
  {Midorikawa}}]{Hondaflux}%
  \BibitemOpen
  \bibfield  {author} {\bibinfo {author} {\bibfnamefont {M.}~\bibnamefont
  {Honda}}, \bibinfo {author} {\bibfnamefont {T.}~\bibnamefont {Kajita}},
  \bibinfo {author} {\bibfnamefont {K.}~\bibnamefont {Kasahara}},\ and\
  \bibinfo {author} {\bibfnamefont {S.}~\bibnamefont {Midorikawa}},\ }\bibfield
   {title} {\bibinfo {title} {Improvement of low energy atmospheric neutrino
  flux calculation using the jam nuclear interaction model},\ }\href
  {https://doi.org/10.1103/PhysRevD.83.123001} {\bibfield  {journal} {\bibinfo
  {journal} {Phys. Rev. D}\ }\textbf {\bibinfo {volume} {83}},\ \bibinfo
  {pages} {123001} (\bibinfo {year} {2011})}\BibitemShut {NoStop}%
\bibitem [{\citenamefont {et~al.}(1994)}]{GEANT3}%
  \BibitemOpen
  \bibfield  {author} {\bibinfo {author} {\bibfnamefont {R.~B.}\ \bibnamefont
  {et~al.}},\ }\bibfield  {title} {\bibinfo {title} {Geant detector description
  and simulation tool},\ }\href
  {https://cds.cern.ch/record/1082634/files/geantall_CERN-W5013.pdf} {\bibfield
   {journal} {\bibinfo  {journal} {CERN Report}\ }\textbf {\bibinfo {volume}
  {CERN-W5013}} (\bibinfo {year} {1994})}\BibitemShut {NoStop}%
\bibitem [{\citenamefont {Shiozawa}(1999)}]{Shiozawa1999}%
  \BibitemOpen
  \bibfield  {author} {\bibinfo {author} {\bibfnamefont {M.}~\bibnamefont
  {Shiozawa}},\ }\bibfield  {title} {\bibinfo {title} {Reconstruction
  algorithms in the super-kamiokande large water cherenkov detector},\ }\href
  {https://doi.org/10.1016/S0168-9002(99)00359-9} {\bibfield  {journal}
  {\bibinfo  {journal} {Nucl. Instrum. Methods Phys. Res., Sect. A}\ }\textbf
  {\bibinfo {volume} {433}},\ \bibinfo {pages} {240} (\bibinfo {year}
  {1999})}\BibitemShut {NoStop}%
\bibitem [{\citenamefont {Amsler}\ \emph {et~al.}(2008)\citenamefont {Amsler}
  \emph {et~al.}}]{Amsler2008}%
  \BibitemOpen
  \bibfield  {author} {\bibinfo {author} {\bibfnamefont {C.}~\bibnamefont
  {Amsler}} \emph {et~al.} (\bibinfo {collaboration} {Particle Data Group}),\
  }\bibfield  {title} {\bibinfo {title} {Review of particle physics},\ }\href
  {https://doi.org/10.1016/j.physletb.2008.07.018} {\bibfield  {journal}
  {\bibinfo  {journal} {Phys. Lett. B}\ }\textbf {\bibinfo {volume} {667}},\
  \bibinfo {pages} {1} (\bibinfo {year} {2008})}\BibitemShut {NoStop}%
\bibitem [{\citenamefont {Roe}\ and\ \citenamefont
  {Woodroofe}(2000)}]{Roe2000}%
  \BibitemOpen
  \bibfield  {author} {\bibinfo {author} {\bibfnamefont {B.~P.}\ \bibnamefont
  {Roe}}\ and\ \bibinfo {author} {\bibfnamefont {M.~B.}\ \bibnamefont
  {Woodroofe}},\ }\bibfield  {title} {\bibinfo {title} {Setting confidence
  belts},\ }\href {https://doi.org/10.1103/PhysRevD.63.013009} {\bibfield
  {journal} {\bibinfo  {journal} {Phys. Rev. D}\ }\textbf {\bibinfo {volume}
  {63}},\ \bibinfo {pages} {013009} (\bibinfo {year} {2000})}\BibitemShut
  {NoStop}%
\bibitem [{\citenamefont {Matsumoto}\ \emph {et~al.}(2022)\citenamefont
  {Matsumoto} \emph {et~al.}}]{Matsumoto2022}%
  \BibitemOpen
  \bibfield  {author} {\bibinfo {author} {\bibfnamefont {R.}~\bibnamefont
  {Matsumoto}} \emph {et~al.} (\bibinfo {collaboration} {Super-Kamiokande
  Collaboration}),\ }\bibfield  {title} {\bibinfo {title} {Search for proton
  decay via p→ $\mu$+ k 0 in 0.37 megaton-years exposure of
  super-kamiokande},\ }\href {https://doi.org/10.1103/PhysRevD.106.072003}
  {\bibfield  {journal} {\bibinfo  {journal} {Phys. Rev. D}\ }\textbf {\bibinfo
  {volume} {106}},\ \bibinfo {pages} {072003} (\bibinfo {year}
  {2022})}\BibitemShut {NoStop}%
\end{thebibliography}%
